\documentclass[aps,prl,reprint,groupedaddress]{revtex4-2}

\usepackage{amsmath}
\usepackage{newtxtext}
\usepackage[varvw]{newtxmath}
\usepackage{comment}

\usepackage{graphicx}
\usepackage{booktabs,dcolumn}
\usepackage{siunitx}
\usepackage{epstopdf}
\usepackage{color}
\usepackage[colorlinks=true,allcolors=blue,bookmarks=true,bookmarksnumbered=true,breaklinks=true]{hyperref}

\begin{document}

\title{Measurement of the Differential Static Scalar Polarizability of the \texorpdfstring{$^\mathbf{88}$Sr$^{+}$}{88Sr+} Clock Transition}

\author{T.~Lindvall}
\email[]{thomas.lindvall@vtt.fi}
\author{K.~J.~Hanhijärvi}
\author{T.~Fordell}
\author{A.~E.~Wallin}
\affiliation{VTT Technical Research Centre of Finland Ltd, National Metrology Institute VTT MIKES, P.O.\ Box 1000, FI-02044 VTT, Finland
}

\date{July 3, 2025}

\begin{abstract}
We report on a precision measurement of the differential static scalar polarizability $\Delta\alpha_0$ of the ${}^{2}\!S_{1/2} \rightarrow {}^{2}\!D_{5/2}$ optical clock transition in the $^{88}$Sr$^{+}$ ion. The polarizability was determined from the `magic' ion-trap drive frequency where the micromotion-induced second-order Doppler and quadratic Stark shifts cancel, using a single clock in an interleaved scheme by switching between minimized and large micromotion.
By measuring at different Mathieu $q_z$ parameters, $\Delta\alpha_0$ can be obtained without prior knowledge of the angle between the rf electric field and the trap axis, which would otherwise dominate the systematic uncertainty.
For validation, measurements were carried out at different micromotion levels and by displacing the ion in opposite directions. The results show excellent consistency and our value, $\Delta\alpha_0 = -4.8314(20)\times 10^{-40}\;\mathrm{J\, m^2/V^2} = -29.303(12)\;\mathrm{au}$, reduces the uncertainty by a factor of 3.5 compared to a previous measurement, while showing a discrepancy of $5 \sigma$. Our measurement reduces the polarizability-related uncertainty of the $^{88}$Sr$^{+}$ clock
to $2.2\times 10^{-19}$ for a blackbody radiation temperature of 295\;K---a significant step towards total uncertainties ${<}1\times10^{-18}$. Using existing polarizability transfer schemes, the result can reduce the uncertainty also for other ion species.

{To appear in Phys.~Rev.~Lett., DOI: \href{https://doi.org/10.1103/52by-28mr}{10.1103/52by-28mr}.}
\end{abstract}

\maketitle

Optical atomic clocks \cite{Ludlow2015b} with fractional frequency uncertainties around \num{1e-18} are sensitive probes for new physics \cite{Safronova2018b}, enable applications such as relativistic geodesy \cite{Mehlstaubler2018a}, and are expected to lead to a redefinition of the second in the International System of Units (SI) \cite{Dimarcq2024a}.
In many optical clocks, the frequency shift due to the ambient blackbody radiation (BBR) is the dominating source of uncertainty. The atomic response to BBR is determined by the differential static scalar polarizability (DSSP) $\Delta\alpha_0$ of the clock transition, with a small dynamic correction from the variation in polarizability over the BBR spectrum.

For ion clocks, theoretical DSSP values have relative uncertainties around 1--4\% for monovalent ions and worse for more complex ones \cite{Safronova2012b}. Considerable efforts have therefore been put into experimental assessment of DSSPs.
Comparing room-temperature and cryogenic clocks is a very fundamental method, but has so far achieved only a 2.6\% uncertainty \cite{Huang2022a}. Measuring the ac Stark shift induced by (near)-infrared laser light has been used for different ion species \cite{Huntemann2016a,Baynham2018b,Arnold2018a,Arnold2019a,Brewer2019a}, but the uncertainty is limited to ${\gtrsim}2\%$ due to the need to determine the absolute light intensity at the ion. This can be avoided by instead measuring magic wavelengths, where the differential dynamic polarizability is zero, but extrapolation to dc limits the uncertainty to ${\gtrsim}0.5\%$ \cite{Barrett2019a,Huang2024a,Barrett2025a}.

For ions whose clock transition has a negative DSSP, the micromotion-induced second-order Doppler (time dilation) and quadratic Stark shifts cancel at a particular `magic' trap drive frequency. This has been used to measure the DSSP in $^{88}$Sr$^+$ \cite{Dube2014a} and $^{40}$Ca$^+$~\cite{Huang2019a} with uncertainties of $1.5\times 10^{-3}$ and $2.9\times 10^{-4}$, respectively. This makes it by far the most accurate method to determine DSSP values and  different methods to transfer the accuracy to ions with positive DSSP have therefore been proposed and demonstrated~\cite{Barrett2019a,Wolf2024a,Wei2024a}.

We have measured the differential polarizability $\Delta\alpha_0$ of the $^{88}$Sr$^{+}$ 
${}^{2}\!S_{1/2} \rightarrow {}^{2}\!D_{5/2}$ clock transition, a secondary representation of the second, using a single ion in an interleaved measurement, switching between minimized and large applied excess micromotion (EMM). We will refer to these two servo locks as the reference and high-EMM clocks, respectively.
This method has the benefit that most frequency shifts, except for the micromotion shifts under study, are common mode to the two clocks. 
We show that by measuring the magic frequency for different Mathieu $q_z$ parameters, the DSSP can be obtained without  knowledge of the angle between the rf electric field and the trap axis.
Our result reduces the uncertainty by a factor of 3.5 compared to \cite{Dube2014a}, while showing a discrepancy of $5\sigma$, corresponding to a fractional frequency correction of \num{-4e-18} at room temperature.

The international metrology community has prepared a roadmap with mandatory criteria that must be fulfilled before a redefinition of the SI second can take place \cite{Dimarcq2024a}. 
Criterion I.1 on uncertainty budgets requires multiple optical clocks with systematic uncertainties ${\lesssim}\num{2e-18}$, while  
criterion I.2 on validation via frequency ratios sets a target uncertainty ${\lesssim}\num{5e-18}$ for clock comparisons. The \num{-4e-18} correction corresponding to the observed polarizability discrepancy is very significant compared to these target uncertainties. 

Since most optical clocks are operated under similar conditions, same-transition clock comparisons are generally not able to detect errors in common atomic parameters, and independent measurements of the parameters are therefore crucial. This is true in particular for the BBR shift and DSSP, as most clocks operate close to room temperature.
Our result will thus benefit the redefinition work, the wide range of $^{88}$Sr$^{+}$ clocks being developed \cite{Akerman2025a,Loh2025a,Spampinato2024a,Steinel2023a,Dube2021a}, and, via polarizability transfer, clocks based on ions with positive DSSP such as $^{171}$Yb$^+$. 
In addition to $^{88}$Sr$^+$ and $^{40}$Ca$^+$, our method can be applied to $^{138}$Ba$^+$ \cite{Arnold2020a}, $^{226}$Ra$^+$ \cite{Holliman2022a}, and the secondary transition in $^{176}$Lu$^+$ \cite{Arnold2024b}.

\begin{figure}[tb]
\includegraphics[width=1\columnwidth]{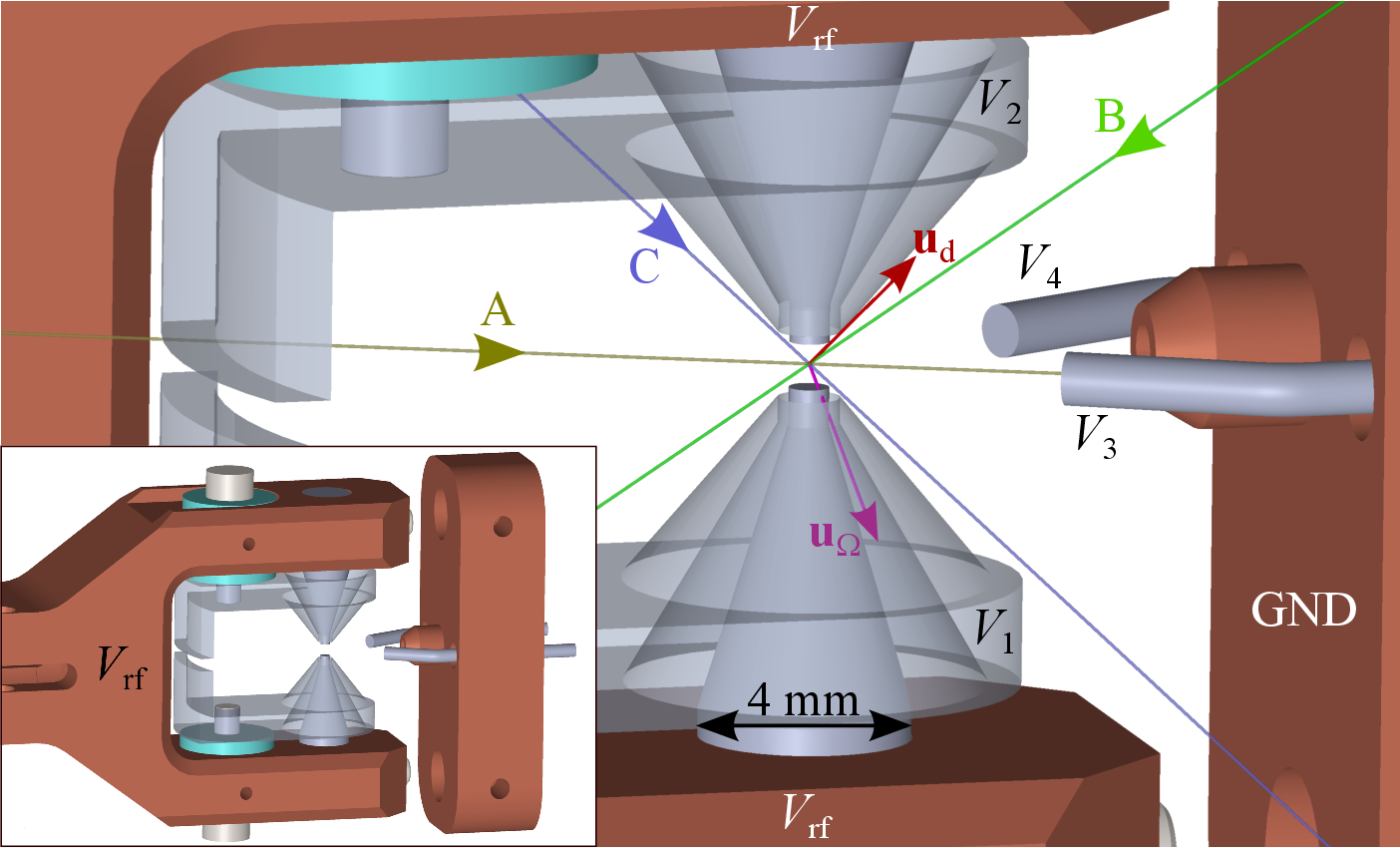}%
\caption{
Trap assembly (inset) and laser-beam geometry.
B is the main cooling and probe beam, while A is the state-preparation beam. All beams A--C are used for micromotion  measurements. For high excess micromotion, a combination of dc bias voltages $V_1$--$V_4$ is applied that causes micromotion along $\mathbf{u}_\Omega$, orthogonal to the A and B beams, while displacing the ion along $\mathbf{u}_\mathrm{d}$, at a $6$--$7^\circ$ angle to the B beam. $V_\mathrm{rf}$---trap rf voltage, GND---ground.
\label{fig:geometry}}
\end{figure}

\emph{Experimental setup}---The endcap ion trap \cite{Lindvall2022a} and laser-beam geometry is shown in Fig.~\ref{fig:geometry} (for details on the laser setup and clock interrogation, see End Matter).
EMM is measured and minimized using photon correlation  \cite{Keller2015a} with three
noncoplanar 422-nm cooling laser beams (A--C in Fig.~\ref{fig:geometry}).
After minimization, the rms rf electric field is $E_\text{rms} \lesssim10$\;V/m.

To apply a large, well-defined EMM, we iteratively determined a combination of dc bias voltages that minimizes the photon-correlation contrast
for the cooling/probe (B) and state-preparation (A) beams, while causing a large contrast for the third beam (C).
Since the relation between the applied dc electric field and the resulting rf field depends on the Mathieu $q_i$ and $a_i$ parameters ($i=x,y,z$ labels the trap axes), proportional to the rf and dc electric field gradients, respectively, a slightly different voltage combination had to be found for each set of parameters.
The dc fields applied for the high-EMM clock were 0.47--1.1\;kV/m, resulting in rms rf fields $E_\text{rms} = 4.3\ldots 6.8$ \;kV/m.
The rf field is orthogonal to the cooling/probe beam (B), but since the $q_i$ parameters have opposite signs in the axial and radial directions, the ion displacement is nearly along the beam (at a $6$--$7^\circ$ angle), see Fig.~\ref{fig:geometry}. Thus even the maximum ion  displacement, $8$\;\si{\micro\m}, causes no significant change in the cooling/probe beam intensity at the ion.
The measurements utilized the full tuning range, 14.22--14.61\;MHz, of the helical resonator, see End Matter for details.

\emph{Theory}---The scalar shift caused by EMM is the sum of the second-order Doppler shift, $\Delta\nu_{\mathrm{D}2} = -(\nu_0/2) \langle v^2\rangle/c^2$, and the quadratic Stark shift, $\Delta\nu_\mathrm{QS} = -\Delta\alpha_0/(2 h) \langle E^2 \rangle$. Here, $\nu_0$ is the transition frequency, $c$ is the speed of light, and $h$ is the Planck constant. From the solution of the Mathieu equation for EMM 
\footnote{See Supplemental Material [url] for derivation of equations and treatment of trap anharmonicity, which includes Refs.~\cite{Schrama1993a,Bentine2020a}.},
the spectrum of the ion's velocity $v$ relates to that of the rf electric field $E$ by $\langle v^2(n\Omega)\rangle = (e/m\Omega)^2 \langle E^2(n\Omega)\rangle/n^2$, where $\Omega$ is the trap drive frequency, $e$ is the elementary charge, $m$ is the ion mass, and $n =1,2,3,\ldots\,$.
This gives the total scalar shift
\begin{equation}\label{eq:dnus}
\Delta\nu_\mathrm{s} = -\frac{\nu_0}{2} \sum_{i=x,y,z} \sum_{n=1}^\infty
\left[ \frac{\Delta\alpha_0}{h \nu_0} + \frac{1}{n^2} \left(\frac{e}{mc \Omega}\right)^2 \right]
\langle E_i^2(n\Omega)\rangle.
\end{equation}
To lowest order in $n$, Eq.~\eqref{eq:dnus} has a zero crossing at $\Omega_0^0 = e/(m c)\sqrt{-h\nu_0/\Delta\alpha_0}$ for an ion with $\Delta\alpha_0<0$.
In previous work \cite{Dube2014a,Huang2019a}, terms up to $n=2$ were included, but for the highest $q_z$ values used here, the third-order term is not negligible and is thus included. Then, from the Mathieu solution \cite{Note1},  
the ratios of the field intensities at $2\Omega$ and $3\Omega$ relative to that at $\Omega$ are given by
\begin{subequations} \label{eq:rhoj}
\begin{eqnarray} 
  \rho_{2,i} = \frac{\langle E_i^2(2\Omega)\rangle}{\langle E_i^2(\Omega)\rangle} =
  \left(\frac{4 q_i}{a_i - 16 - q_i^2/(a_i - 36)}\right)^2, \label{eq:rho2} \\
  \rho_{3,i} = \frac{\langle E_i^2(3\Omega)\rangle}{\langle E_i^2(\Omega)\rangle} =
  \left(\frac{9 q_i^2}{(a_i - 36)(a_i - 16) - q_i^2}\right)^2.
\end{eqnarray}
\end{subequations}

The first harmonic of the rf field, $\mathbf{E}(\Omega)$,  has a well-defined direction $\mathbf{u}_\Omega = \mathbf{E}(\Omega)/|\mathbf{E}(\Omega)|$ orthogonal to the beams A and B, see Fig.~\ref{fig:geometry}, as we minimize their photon-correlation contrasts at $\Omega$.
Noting that $\langle E_i^2(\Omega)\rangle = u_{\Omega,i}^2 \langle E^2(\Omega)\rangle$ and defining the mean ratios $\bar{\rho}_j = \sum_i u_{\Omega,i}^2 \rho_{j,i}$, we can rewrite Eq.~\eqref{eq:dnus} as
\begin{equation}\label{eq:dnus2}
\Delta\nu_\mathrm{s} \approx -\frac{\Delta\alpha_0}{2 h}
\left[ 1 - \left(\frac{\Omega_0^0}{\Omega}\right)^2 \frac{1 + \bar{\rho}_2/4 + \bar{\rho}_3/9}{1 + \bar{\rho}_2 + \bar{\rho}_3} \right] \langle E^2\rangle,
\end{equation}
where $\langle E^2\rangle \approx (1 + \bar{\rho}_2 + \bar{\rho}_3) \langle E^2(\Omega)\rangle$ is the total mean-square rf field. For a constant applied dc field, $\langle E^2\rangle$ depends on the Mathieu parameters but not on $\Omega$.

In \cite{Lindvall2022a}, the rf gradient of our trap was found to be radially symmetric within experimental uncertainty. 
As the $a_i$ parameters are very small, $|a_i|\ll 1$, we can  initially neglect them and finally evaluate the small correction they cause.
This justifies neglecting the radial asymmetry and writing 
$\bar{\rho}_j = \rho_{j,z} \cos^2{\theta_\Omega} + \rho_{j,r} \sin^2{\theta_\Omega}$, where $\theta_\Omega$ is the angle between $\mathbf{u}_\Omega$ and the trap $z$ axis (vertical in Fig.~\ref{fig:geometry}) \footnote{Note that in \cite{Dube2014a,Huang2019a}, the angle $\beta$ is the angle between the total rf electric field $\mathbf{E}$ and the trap $z$ axis.}.

Defining the zero crossing of Eq.~\eqref{eq:dnus2} as
\begin{equation}\label{eq:Omega0}
\Omega_0 = \Omega_0(\bar{\rho}_2, \bar{\rho}_3)
= \Omega_0^0 \sqrt{\frac{1 + \bar{\rho}_2/4 + \bar{\rho}_3/9}{1 + \bar{\rho}_2 + \bar{\rho}_3}},
\end{equation}
the expression in square brackets in \eqref{eq:dnus2} becomes $1 - \left( \Omega_0/\Omega \right)^2$.
This function is not well suited for fitting to data over a narrow frequency range,
so we use a second-order series expansion in $\delta/\Omega_0$, where $\delta = \Omega - \Omega_0$ is the detuning, which yields
\begin{equation}\label{eq:dnus3}
\Delta\nu_\mathrm{s} \approx k
\left[ \frac{\delta}{\Omega_0}  - \frac{3}{2} \left( \frac{\delta}{\Omega_0} \right)^2 \right],
\end{equation}
where $k = (-\Delta\alpha_0/h) \langle E^2\rangle$. For our parameters, the maximum value of the second-order term is of the same order as the statistical uncertainties, while the third-order term is negligible. Since we keep $q_z$ constant when changing $\Omega$ (and $|a_i|\ll 1$), $\langle E^2\rangle$ is constant and Eq.~\eqref{eq:dnus3} has only two fit parameters, the zero crossing $\Omega_0$ and the slope at the zero crossing, $k$. By contrast, in \cite{Dube2014a,Huang2019a} the ion secular frequencies were kept constant, which causes $\langle E^2\rangle$ to have a complicated dependence on $\Omega$ and necessitates using a physically less meaningful fit function with three free parameters.

\emph{Differential frequency shifts}---Although most frequency shifts are common mode to the two clocks, we must carefully evaluate those that may differ.
A large first-order Doppler shift would occur if the ion were probed during the displacement of the ion's equilibrium position following the switching of the voltages.
This was studied, see End Matter for details, and we estimate that a dwell time of 300\;ms between the switching and probing gives negligible shifts of ${<}30$\;\si{\micro\Hz} for all measurements, which we include as uncertainties.

Since the trap has only two bias electrodes in the radial plane, changing the electric field at the position of the ion also changes its gradient \cite{Lindvall2022a}, causing a small change in the electric quadrupole shift (EQS). This does not affect the scalar shift studied here due to the efficient EQS cancellation scheme \cite{Dube2005a,Lindvall2022a}, but it prevents us from determining the tensor polarizability, as the EQS and the tensor Stark shift, both tensor shifts, cannot be reliably distinguished.

We measured the ion temperature and heating rate without and with applied EMM for different $q_z$ and $\Omega$ using carrier Rabi flopping.
In most cases, the difference in estimated mean ion temperature was small, ${\lesssim}0.3$\;mK. However, for a few secular frequencies we observed an increased heating rate. These resonances were narrower than the small change in secular frequencies due to the change in dc gradient with applied EMM and lead to a temperature difference of ${\approx}2$\;mK magnitude for two measurements (out of 24). These were corrected for, and a conservative uncertainty of 1\;mHz (1.4\;mK) was assigned to the differential thermal second-order Doppler shift of all measurements.

\emph{Measurements}---Clock interrogation times of 124\;ms or 137\;ms were used. The frequency difference between the high-EMM and reference clocks averaged down as ${\approx}7\times10^{-15} \tau^{-1/2}$, resulting in a statistical uncertainty around $10$\;mHz after one day of averaging for each combination of parameters (rf frequency, $q_z$, and ion displacement).
The secular frequencies were measured before and after each measurement to determine the Mathieu parameters, see End Matter for details.

The maximum ion displacement was ultimately limited by the range of the bias voltage supply, but also by the finite field of view of the fluorescence detection. The measurements at $q_z = 0.34\ldots 0.46$ were carried out with small $a_i$ parameters of $(-0.0014, 0.0004, 0.0010)$ used to minimize the EQS \cite{Lindvall2022a}. For even lower $q_z$, we were not able to achieve high EMM shifts without a significantly increased risk of ion loss when switching the voltages. For the highest $q_z$ values of 0.60 and 0.71, the trap was operated with increased $a_i$ parameters of $(0.0047, 0.0069, -0.0117)$ in order to achieve a sufficient ion displacement.

\begin{figure}[tb]
\includegraphics[width=1\columnwidth]{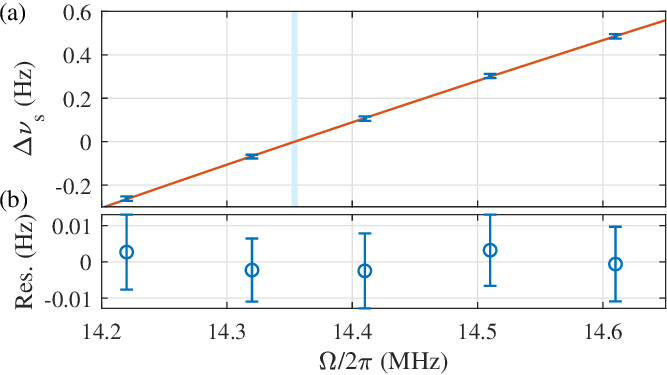}%
\caption{(a) Measured micromotion-induced scalar frequency shifts with statistical uncertainties (error bars) as function of the rf drive frequency for $q_z = 0.37$ and an applied dc field of $0.65$\;kV/m.
The line is a fit of Eq.~\eqref{eq:dnus3} and the shaded area indicates the uncertainty of the zero crossing frequency ($\pm1\sigma$). (b) Residuals of fit in (a).
\label{fig:shifts}}
\end{figure}

\emph{Analysis and results}---Figure~\ref{fig:shifts} shows an example of measured EMM-induced shifts fitted with Eq.~\eqref{eq:dnus3}.
For the remaining six measurements, 3--4 points were measured. In total, we collected 23.4~days of data for 24 points. 

The measured zero crossing frequency depends on the angle $\theta_\Omega$ between $\mathbf{u}(\Omega)$ and the trap axis.
Based on the laser beam directions, the angle between $\mathbf{u}(\Omega)$  and the vertical \emph{chamber} axis is $32(1)^\circ$,
but the direction of the trap axis is not known as accurately. In previous work \cite{Dube2014a,Huang2019a}, the angle was determined from the tensor shift measured with the magnetic field aligned with the trap axis. In addition to relying on  theoretical tensor polarizability values, this method requires estimates of the direction of the trap axis and the accuracy of the magnetic field direction. It is also unclear whether the EQS was taken into account, and we noted that even a small change in it prevented us from determining the tensor polarizability.

\begin{figure}[tb]
\includegraphics[width=1\columnwidth]{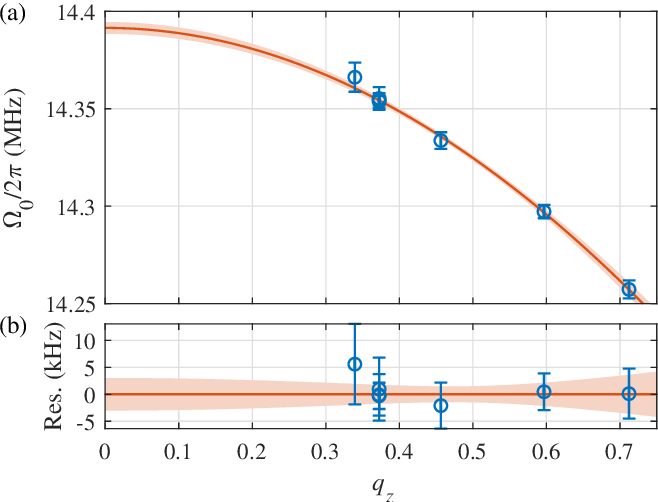}%
\caption{(a) Measured zero crossing frequencies with statistical uncertainties (error bars) as function of $q_z$. Each point is the result of a fit like in Fig.~\ref{fig:shifts}. The uncertainties of the $q_z$ values are ${\lesssim}\num{3e-4}$ and the corresponding error bars are not visible. 
The line is a fit of Eq.~\eqref{eq:Omega0} and the shaded area the corresponding uncertainty ($\pm1 \sigma$). (b) Residuals of fit in (a).
\label{fig:qz_fit}}
\end{figure}

In our case, the angle $\theta_\Omega$ would dominate the  systematic uncertainty. Instead of evaluating its value and uncertainty using multiple assumptions, we measure $\Omega_0$ over a wide range of $q_z$ parameters, which allows determining the DSSP without knowledge of the angle. Neglecting the small $a_i$ parameters, the zero crossing, Eq.~\eqref{eq:Omega0}, depends only on $\Omega_0^0$, $\theta_\Omega$ and $q_z$, and the first two parameters can be obtained from a fit of $\Omega_0$ as a function of $q_z$, see Fig.~\ref{fig:qz_fit}.
This gives $\Omega_0^0/2\pi = 14.3915(30)$\;MHz and $\theta_\Omega = 31.2(3.1)^\circ$, in agreement with the nominal angle given above.

To demonstrate the robustness of the $q_z$ fitting method, we repeat the fitting using (i) the second-order Mathieu solution, which yields $\Omega_0^0/2\pi = 14.3916(30)$\;MHz and $\theta_\Omega = 30.9(3.1)^\circ$, and (ii) a series expansion to order $q_z^2$, 
which yields $\Omega_0^0/2\pi = 14.3909(30)$\;MHz and $\theta_\Omega = 32.5(2.9)^\circ$ \cite{Note1}.
This shows that the angle $\theta_\Omega$ adapts to errors in the fit function, while the effect on $\Omega_0^0$, and thus $\Delta\alpha_0$, is relatively small even for significant errors or approximations.

From the measured residual photon-correlation contrasts with high EMM, we estimate that the applied micromotion is orthogonal to the A and B beams with an uncertainty  ${\lesssim} 0.9^\circ$. However, any variations in the direction are accounted for by the fit uncertainty.
The systematic uncertainty contributions from $q_z$, the first-order Doppler shift, and the ion temperature difference were evaluated using a Monte-Carlo simulation.
The effect of reference-clock residual EMM was found to be negligible, see End Matter for details.
To take into account the so far neglected $a_i$ parameters, we correct the measured zero-crossing frequencies using the square root in Eq.~\eqref{eq:Omega0} evaluated with and without $a_i$ parameters \footnote{This requires the azimuthal angle of the rf direction, $\varphi_\Omega$, whose uncertainty of ${\approx}5^\circ$ has a completely negligible effect.}. Fitting the third-order solution gives $\Omega_0^0/2\pi = 14.3916(30)$\;MHz and $\theta_\Omega = 31.1(3.1)^\circ$. The uncertainty contribution from the $a_i$ parameters is conservatively taken as the difference in $\Delta\alpha_0$ between included and neglected $a_i$.

\begin{table}[b]
\centering
  \caption{
    $\Delta\alpha_0$ uncertainty budget.
   \label{tab:u-budget}}
   \begin{ruledtabular}
  \begin{tabular}{l S}
Contribution & {Uncertainty ($10^{-43}\;\mathrm{J m^2/V^2}$)}  \\
\midrule
Statistical/fitting & 2.0 \\
$q_z$ parameter     & 0.0072 \\
Ion temperature     & 0.22 \\
First-order Doppler (dwell time)  & 0.0046 \\
Reference-clock residual EMM & 0.0003 \\
$a_i$ parameters    & 0.053 \\
Trap anharmonicity  & 0.0020 \\
Total               & 2.0 \\
  \end{tabular}
  \end{ruledtabular}
\end{table}

We also considered the effect of anharmonic terms in the trap potential and found that the $q_z$ fitting reduces the corresponding DSSP error by one order compared to a measurement at a single $q_z$ value \cite{Note1}. 
This completes the uncertainty budget, summarized in Table~\ref{tab:u-budget}, and gives the final DSSP value $\Delta\alpha_0 = -4.8314(20)\times 10^{-40}\;\mathrm{J\, m^2/V^2} = -29.303(12)\;\mathrm{au}$.

\begin{figure}[tb]
\includegraphics[width=1\columnwidth]{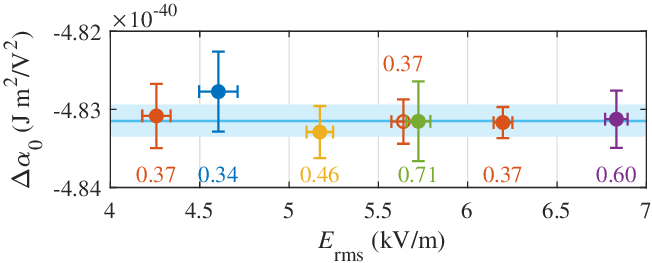}%
\caption{$\Delta\alpha_0$ values from the individual measurements as function of the rms rf field with $q_z$ values below/above. For the open circle, the ion was displaced in the opposite direction. The line and shaded area are the $\Delta\alpha_0$ value and uncertainty from the $q_z$ fitting.
\label{fig:DSSP_ind}}
\end{figure}

For $q_z = 0.37$, we measured the zero crossing with three different ion displacements, one of them in the opposite direction ($-\mathbf{u}_\mathrm{d}$ in Fig.~\ref{fig:geometry}).
As seen from Fig.~\ref{fig:qz_fit}, these agree well.
With the angle $\theta_\Omega$ known, we can also solve the DSSP separately from each measurement. These results are shown in Fig.~\ref{fig:DSSP_ind} as a function of $E_\text{rms}$, solved from the slope $k$ in Eq.~\eqref{eq:dnus3}. Again, no trends with $E_\text{rms}$, $q_z$, or the direction of displacement are seen. For simplicity, we have here neglected the correlation between the fitted $\theta_\Omega$ and the individual measurements.

\begin{figure}[b]
\includegraphics[width=1\columnwidth]{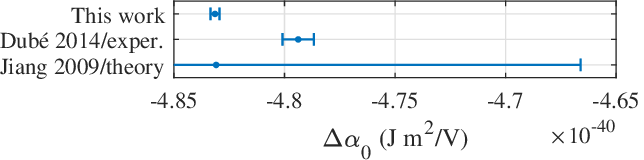}%
\caption{ $\Delta\alpha_0$ measured in this work compared to previous experimental \cite{Dube2014a} and theoretical \cite{Jiang2009a} values.
\label{fig:comp}}
\end{figure}

\emph{Discussion}---As seen from Fig.~\ref{fig:comp}, our $\Delta\alpha_0$ value shows a discrepancy of $5\sigma$ with the experimental result in \cite{Dube2014a}. The reason for this  is unknown. E.g., an error in the angle between the rf field and the trap axis in \cite{Dube2014a} could only explain about half of the difference. However, we note that the high-EMM clock used in \cite{Dube2014a} was not very well characterized \cite{Madej2004a}. With only a single laser beam, there was limited knowledge about the ion temperature and EMM. As the EMM could not be minimized, there was also no reference measurement where the two clocks agreed, but the measurement relied solely on the estimated systematic shifts. With measurements at a single EMM level, any undiagnosed frequency offset between the two clocks  would cause an offset in the DSSP value. By contrast, since we get consistent results at different EMM levels, a bias in our DSSP value could only be caused by an unknown shift proportional to $\langle E^2\rangle$.
Our value is close to the theoretical DSSP value  \cite{Jiang2009a} but has an 80 times smaller uncertainty and can therefore provide a benchmark for future atomic structure calculations.

Our $\Delta\alpha_0$ value reduces the polarizability-related uncertainty of the $^{88}$Sr$^{+}$ clock to $2.2\times 10^{-19}$ for a BBR temperature of 295\;K. Reducing the BBR-field--related uncertainty to the same level would require a BBR temperature uncertainty of 31\;mK.
Still, since our uncertainty is dominated by statistics, the $q_z$ fitting method has potential to  further reduce the $\Delta\alpha_0$ uncertainty to a level limited by the repeatability of the applied micromotion and the remaining systematics, see End Matter for details.

Our $\Delta\alpha_0$ value implies a frequency correction of $-4\times 10^{-18}$ for  room-temperature ${}^{88}$Sr$^+$ clocks that have used the value from \cite{Dube2014a}. Among published  results, this correction is significant compared to the total uncertainty only for the ${}^{88}$Sr$^+$/${}^{171}$Yb$^+$(E3) optical frequency ratio in \cite{Steinel2023a}.

\emph{Summary}---We have measured the differential polarizability $\Delta\alpha_0$ of the $^{88}$Sr$^{+}$ clock transition by utilizing the `magic' trap drive frequency where the micromotion shifts cancel. The measurements were carried out using a single ion trap in an interleaved scheme, resulting in common-mode rejection of most other  shifts. We have showed that by measuring at different Mathieu $q_z$ values, $\Delta\alpha_0$ can be obtained without knowledge of the angle between the rf  field and the trap axis.
Our measurements show excellent consistency and the fractional uncertainty of \num{4.1e-4}, a factor of 3.5 lower than the previous value, reduces the polarizability-related uncertainty of the $^{88}$Sr$^{+}$ clock to $2.2\times 10^{-19}$ for a BBR temperature of 295\;K. Using existing polarizability transfer schemes, the improved uncertainty can benefit also other ion species.

\begin{acknowledgments}
\emph{Acknowledgments}---We thank Pierre Dub\'e for helpful discussions.
This work has been supported by the projects 23FUN03 HIOC and 22IEM01 TOCK, which have received funding from the European Partnership on Metrology, co-financed from the European Union’s Horizon Europe Research and Innovation Programme and by the Participating States.
The work is also part of the Academy of Finland Flagship Programme `Photonics Research and Innovation' (PREIN, decision 320168).
\end{acknowledgments}

\emph{Data availability}---The data that support the findings of this article are openly available \cite{Note4}.

%

\onecolumngrid
\section{End Matter}
\twocolumngrid

\emph{Laser system and interrogation sequence}---The ion is Doppler cooled using a commercial external-cavity diode laser at 422\;nm, frequency stabilized to a nearby rubidium line~\cite{Shiner2007a} using modulation transfer spectroscopy.
For repumping at 1092\;nm and clock-state clearout at 1033\;nm, unpolarized amplified-spontaneous-emission (ASE) light sources \cite{Lindvall2013a,Fordell2015a} are used. After cooling, the main cooling beam (B in Fig.~\ref{fig:geometry}) is switched off and a separate 422-nm state-preparation beam (A in Fig.~\ref{fig:geometry}),  propagating along the magnetic field direction and whose polarization can be controlled using a liquid-crystal phase retarder, is briefly switched on with $\sigma^+$ or $\sigma^-$ polarization to transfer the ion into the desired ground-state sublevel. The two beams are controlled using independent double-pass acousto-optic modulators (AOMs), while a common mechanical shutter ensures that the 422-nm light is fully extinguished during clock interrogation. The clock transition is probed using a commercial 1348-nm external-cavity diode laser, which is frequency-doubled to 674\;nm and stabilized to a 30-cm horizontal ultra-low-expansion (ULE) glass cavity of a design similar to \cite{Hafner2015a}.
After probing, the cooling laser and repumper are switched on for state detection. Then a short 1033-nm clearout pulse is applied to return the ion from the clock state to the cooling cycle, followed by a second state detection to verify fluorescence at 422\;nm. Ion fluorescence is detected using an objective close to the anti-reflection--coated fused silica window of the vacuum system. A beam splitter divides the detected photons between a photomultiplier tube (PMT, photon collection efficiency ${\approx}1/550$) and an sCMOS camera.

\begin{figure}[t]
\includegraphics[width=1\columnwidth]{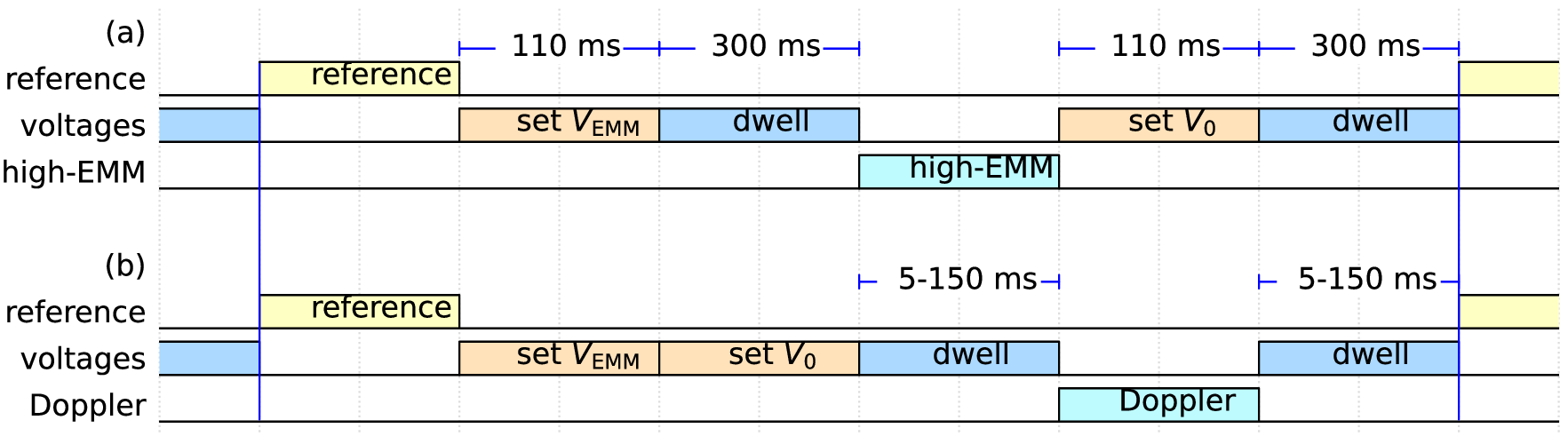}
\caption{(a) Interleaved clock sequence for measuring the DSSP. (b) Interleaved clock sequence for measuring the first-order Doppler shift with variable dwell time. The blue vertical lines indicate the start and end of an interleaved clock cycle, with a total duration of 8--9\;s (depending on the probe time). $V_\mathrm{EMM}$---bias voltages for high EMM, $V_0$---bias voltages for minimized EMM.
\label{fig:interl}}
\end{figure}

Both the reference and high-EMM clocks probed six $S_{1/2} \rightarrow D_{5/2}$ Zeeman components, $\pm1/2 \rightarrow \pm1/2$, $\pm1/2 \rightarrow \pm3/2$, and $\pm1/2 \rightarrow \pm5/2$, in order to cancel the electric quadrupole shift and other tensor shifts \cite{Dube2005a,Lindvall2022a} in addition to the linear Zeeman shift.
Each frequency (red and blue side of six transitions) was probed twice per cycle for both clocks. The order of probing the red and blue sides was  alternated in order to avoid noise-induced servo errors \cite{Lindvall2023a}.
The interleaved clock sequence is schematically shown in Fig.~\ref{fig:interl}(a).
The frequency difference between the high-EMM and reference clocks was evaluated from the frequency difference of the final probe-laser AOM before the ion.

\emph{Trap drive}---To allow accurate and repeatable tuning of the resonance frequency between 14.22\;MHz and 14.61\;MHz, our tunable helical resonator \cite{Lindvall2022a} was equipped with a micrometer tuning screw. We use a phase-continuously tunable
Direct Digital Synthesizer (DDS) rf source (Sinara Urukul/AD9912) with a power amplifier and rf transformer (for galvanic isolation) to deliver rf to the helical resonator input, which allows changing the frequency without ion loss. We adjust $q_z$ coarsely using a digital step attenuator with 0.5-dB resolution and finely, to keep $q_z$ constant when changing the rf frequency, using the 10-bit full-scale current register of the DDS (relative amplitude resolution~$10^{-3}$).

\emph{First-order Doppler shift}---The dc voltage source channels have time constants of ${\approx}30$\;ms, including the low-pass filters at the electric vacuum feedthroughs. We  must therefore introduce a dwell time between the voltage switching and the probe pulse to ensure that the first-order Doppler shift due to the displacement of the ion's equilibrium position is at a negligible level. To determine the required dwell time, the clock sequence in Fig.~\ref{fig:interl}(b) was used to measure the residual Doppler shift with different dwell and probe times.
For technical reasons, the four voltage channels are switched one at a time, with a total duration of ${\approx}110$\;ms.

For the shortest dwell times, the initial Doppler shift is large enough that the first probe pulse after switching does not excite the ion. Due to the alternating red/blue probing (see above), the servo tracking still works well. Even for a short interrogation time of 36.5\;ms, the excitation probability returned to normal already for a dwell time of 50\;ms. Here, the measured line center shift was ${\approx}500$\;mHz and changed sign with the ion displacement direction.  In the ${\geq}50$\;ms dwell-time regime, a model that assumes that an  exponentially decaying instantaneous frequency shift is averaged over each probe pulse gives a good fit. This allows extrapolating to longer dwell and probe times, and we estimate that a dwell time of 300\;ms gives negligible shifts of ${<}30\;$\si{\micro\Hz} for all measurements. The agreement between the $\Omega_0$ (Fig.~\ref{fig:qz_fit}) and $\Delta\alpha_0$ values (Fig.~\ref{fig:DSSP_ind}) measured with opposite ion displacement also confirms that the 300-ms dwell time is sufficient.

\emph{Secular frequencies and Mathieu parameters}---Secular frequencies were measured using the tickler voltage and photon correlation method \cite{Lindvall2022a}. For the highest $q_z$ values of 0.60 and 0.71, the accuracy of the method to determine Mathieu parameters presented in \cite{Lindvall2022a} is not sufficient. The obtained $q_z$ and $a_i$ parameters were therefore iteratively corrected by estimating their error, correcting for it, and estimating the new error. This method is based on the fact that the secular frequencies can be calculated to arbitrary precision from the Mathieu parameters, but not vice versa. After two iterations, the systematic error of the obtained parameters was negligible compared to their stability/repeatability. The latter was estimated as the standard deviation of all values measured for a certain $q_z$ and ion displacement (including all measured rf frequencies).

\emph{Reference-clock EMM}---The EMM of the reference clock was measured and, if required, minimized before and after each measurement. From these, we interpolate the mean rf electric field during each EMM-shift measurement. An rms field ${\lesssim}40\;\mathrm{V/m}$ and a corresponding frequency shift of magnitude ${\lesssim}12$\;\si{\micro\Hz} were found for all measurements.
EMM in the direction of the applied (high) EMM cancels in the differential measurement, while EMM in the orthogonal direction causes a bias in the measured shift. The angle between the estimated reference EMM and the applied EMM varied between $20^\circ$ and $70^\circ$, so some suppression is expected, but the uncertainty contribution from the reference clock EMM
was conservatively estimated by correcting for the reference EMM shifts in full, repeating the fitting, and taking the change in $\Delta\alpha_0$ as the uncertainty. This yielded a negligible contribution as shown in Table~\ref{tab:u-budget}.

\emph{Uncertainty reduction}---As our $q_z$ fitting method reduces the angle $\theta_\Omega$ to a fit parameter, it allows further reduction of the total DSSP uncertainty to a level limited by the reproducibility of the applied EMM and the remaining systematics if the statistical uncertainty can be reduced.
Using two separate clocks instead of an interleaved measurement would reduce the instability of the frequency difference by a factor of ${\approx}1.5$ for our parameters, providing a statistical advantage despite the need for separate reference measurements. A two-clock measurement would also
allow significantly lower $q_z$ values to be used, which would reduce the fitting uncertainty. Finally, even higher EMM levels would allow determining the zero crossing frequency with lower uncertainty.

\onecolumngrid
\clearpage
\newpage

\begin{center}
{\bf \large Supplemental Material: Measurement of the Differential Static Scalar Polarizability of the $^\mathbf{88}$Sr$^{+}$ Clock Transition}\\
\vspace{\baselineskip}
T.~Lindvall,$^*$ K.~J.~Hanhij\"arvi, T.~Fordell and A.~E.~Wallin\\
\emph{\small VTT Technical Research Centre of Finland Ltd, National Metrology Institute VTT MIKES, P.O.\ Box 1000, FI-02044 VTT, Finland}\\
{\small (Dated: July 3, 2025)}\\
\vspace{\baselineskip}
\end{center}
\twocolumngrid

\counterwithin*{figure}{part}
\counterwithin*{table}{part}
\counterwithin*{equation}{part}
\stepcounter{part}
\setcounter{page}{1}
\renewcommand{\theequation}{S\arabic{equation}}
\renewcommand{\thefigure}{S\arabic{figure}}
\renewcommand{\thetable}{S\Roman{table}}
\thispagestyle{empty}

\section{Mathieu solution}

Here, we summarize the solution of the nonhomogeneous Mathieu equation for EMM (see the Supplemental Material of \cite{sDube2014a} for more details) and derive the relations used in the main text. The motion of the ion is the sum of the nonhomogeneous solution derived here and the homogeneous solution for thermal motion, but the latter is not needed for this work.

The electric potential of a spherical Paul trap can be written as \cite{sLindvall2022a}
\begin{equation} \label{eq:phiaq}
\phi(\mathbf{x}) = \frac{m \Omega^2}{8 e} \sum_i \left(a_i - 2q_i \cos{\Omega t} \right) x_i^2,
\end{equation}
where $\mathbf{x} = (x_x, x_y, x_z) = (x, y, z)$, $m$ is the ion mass, $\Omega$ the rf frequency, $e$ the elementary charge, the summation is over the trap axes, and the Mathieu parameters $a_i$ and $q_i$ are proportional to the gradient of the dc and rf electric field, respectively. The total electric field along the trap axis $i$ is then 
\begin{equation} \label{eq:Etot}
E_{\text{tot},i} = -\frac{d \phi}{dx_i} + E_{\mathrm{dc},i} = -\frac{m \Omega^2}{4 e} \left(a_i - 2q_i \cos{\Omega t} \right) x_i + E_{\mathrm{dc},i},
\end{equation}
where $E_{\mathrm{dc},i}$ is the dc field component. The equation of motion, $m d^2x_i/dt^2 = e E_{\text{tot},i}$, can then be written in the form of a nonhomogeneous Mathieu equation, 
\begin{equation} \label{eq:mathieu}
\frac{d^2 x_i}{d\tau^2} + \left(a_i - 2q_i \cos{2\tau}\right) x_i = \frac{4 e}{m\Omega^2}  E_{\mathrm{dc},i},
\end{equation}
where $\tau = \Omega t/2$. The solution for EMM can be written as 
\begin{equation}
x_i(\tau)=\sum_{n=-\infty}^\infty c_{2n,i} \cos{(2n\tau)}.
\end{equation}
Substituting $x_i$ and its second derivative into Eq.~\eqref{eq:mathieu} leads to recurrence relations for the coefficients $c_{2n,i}$
\begin{eqnarray} 
    \left(a_i - 4n^2\right)c_{2n,i} - q_i\left(c_{2(n+1),i} + c_{2(n-1),i}\right) \nonumber \\
    = 
    \begin{cases}
			\frac{4 e}{m\Omega^2}  E_{\mathrm{dc},i}, & \text{$n=0$},\\
            0, & \text{$n \neq 0$}.
		 \end{cases} \label{eq:rec}
\end{eqnarray}
This equation is symmetric with respect to positive and negative $n$, so $c_{2n,i} = c_{-2n,i}$. This gives
\begin{equation} \label{eq:x-sol}
x_i(t)= c_{0,i} + 2\sum_{n=1}^\infty c_{2n,i} \cos{(n\Omega t)},
\end{equation}
and the ion velocity becomes 
\begin{equation} \label{eq:vel}
v_i(t) = \frac{d x_i}{dt}= -2\Omega \sum_{n=1}^\infty n c_{2n,i} \sin{(n\Omega t)}.
\end{equation}
Substituting Eq.~\eqref{eq:x-sol} into \eqref{eq:Etot} and simplifying using the recurrence relations \eqref{eq:rec} gives \cite{sDube2014a}
\begin{equation} \label{eq:Etot2}
E_{\text{tot},i} = -\frac{2m \Omega^2}{e} \sum_{n=1}^\infty n^2 c_{2n,i} \cos{(n\Omega t)}.
\end{equation}
Equations~(\ref{eq:vel}--\ref{eq:Etot2}) give the relation between the spectrum of the ion's velocity and the spectrum of the rf electric field, $\langle v^2(n\Omega)\rangle = (e/m\Omega)^2 \langle E^2(n\Omega)\rangle/n^2$, which is used for Eq.~(1) in the main text. Equation~\eqref{eq:Etot2} also gives the ratio of the harmonic and fundamental field intensities used for Eq.~(2) in the main text,
\begin{equation} \label{eq:rhon}
\rho_{n,i} = \frac{\langle E_i^2(n\Omega)\rangle}{\langle E_i^2(\Omega)\rangle} =
  \left(\frac{n^2 c_{2n,i}}{c_{2,i}}\right)^2.
\end{equation}

To solve the Mathieu equation to order $n$, we set $c_{2(n+1),i}=0$ and apply the recursion relations \eqref{eq:rec}. To third order, the coefficients become
\begin{subequations} \label{eq:c}
\begin{eqnarray} 
  c_{0,i} &=& \frac{\frac{4 e}{m\Omega^2}  E_{\mathrm{dc},i}}{a_i-2\frac{q_i^2}{a_i-4-\frac{q_i^2}{a_i-16-\frac{q_i^2}{a_i-36}}}}, \label{eq:c0} \\
  c_{2,i} &=& \frac{q_i c_{0,i}}{a_i-4-\frac{q_i^2}{a_i-16-\frac{q_i^2}{a_i-36}}}, \\
  c_{4,i} &=& \frac{q_i c_{2,i}}{a_i-16-\frac{q_i^2}{a_i-36}}, \\  
  c_{6,i} &=& \frac{q_i c_{4,i}}{a_i-36}.
\end{eqnarray}
\end{subequations}
We note that the coefficients take the form of continued fractions with the numerical factors in the denominators being $(2n)^2$. Inserting Eqs.~\eqref{eq:c} into \eqref{eq:rhon} gives Eqs.~(2) in the main text:
\begin{subequations} \label{eq:rhojs}
\begin{eqnarray} 
  \rho_{2,i} = \frac{\langle E_i^2(2\Omega)\rangle}{\langle E_i^2(\Omega)\rangle} =
  \left(\frac{4 q_i}{a_i - 16 - q_i^2/(a_i - 36)}\right)^2, \label{eq:rho2s} \\
  \rho_{3,i} = \frac{\langle E_i^2(3\Omega)\rangle}{\langle E_i^2(\Omega)\rangle} =
  \left(\frac{9 q_i^2}{(a_i - 36)(a_i - 16) - q_i^2}\right)^2.
\end{eqnarray}
\end{subequations}

Also the second-order Mathieu solution is used for fitting the measured zero-crossing frequencies. It is obtained by setting $c_{6,i}=0$ and also omitting the terms $-q_i^2/(a-36)$ in Eqs.~(\ref{eq:c}a--c). Specifically, the field intensity ratio becomes
\begin{equation} \label{eq:rho2-2}
      \rho_{2,i} = 
  \left(\frac{4 q_i}{a_i - 16}\right)^2. 
\end{equation}

\section{Series expansion of zero-crossing frequency}

A series expansion of the zero-crossing frequency $\Omega_0$ to second order in $q_z$ is also used in the main text. To this order, we can start with the second-order result \eqref{eq:rho2-2}, which, when neglecting the $a_i$ parameters, becomes $\rho_{2,i} = (q_i/4)^2$. Thus $\rho_{2,r} = \rho_{2,z}/4$ and $\bar{\rho}_2 = \rho_{2,z} \cos^2{\theta_\Omega} + \rho_{2,r} \sin^2{\theta_\Omega} = (1+3 \cos^2{\theta_\Omega}) q_z^2/2^6$. 
Substituting this into Eq.~(4) in the main text and doing a series expansion to order $q_z^2$ gives 
\begin{equation}\label{eq:Omega0s}
\Omega_0 \approx \Omega_0^0 \sqrt{\frac{1 + \bar{\rho}_2/4 }{1 + \bar{\rho}_2 }} \approx \Omega_0^0\left[ 1 - \frac{3}{2^9}q_z^2 (1+3 \cos^2{\theta_\Omega})\right].
\end{equation}

\section{Trap anharmonicity}

Anharmonic terms in the potential require cross terms between the trap axes to fulfill the Laplace equation. As an example, we consider the following potential with small $z^3$ and $z^4$ terms in the rf potential 
\begin{widetext}
\begin{equation} \label{eq:phi2}
\phi(\mathbf{x}) = \frac{m \Omega^2}{8 e} \Bigg\{\sum_i \left(a_i - 2q_i \cos{\Omega t} \right) x_i^2 - 2q_z \cos{\Omega t}    
\bigg[\varepsilon_{3,z} \left(z^3-\frac{3}{2}x^2z -\frac{3}{2}y^2z\right) +\varepsilon_{4,z} \left(z^4 -3x^2z^2 -3y^2z^2 +3x^2y^2\right) \bigg] \Bigg\}.
\end{equation}
\end{widetext}
A possible anharmonicity in the dc potential can be shown to have a negligible effect compared to that of the rf potential and is omitted. The anharmonic coefficients $\varepsilon_{j,i}$ give small corrections for typical ion displacements. The fourth-order term arises in endcap traps~\cite{sSchrama1993a} if the relations between the electrode dimensions are not optimal, whereas the third-order term can result from any symmetry breaking in the trap design or fabrication. 

Ion motion in the potential \eqref{eq:phi2} was numerically simulated in (py)LIon \cite{sBentine2020a}. With a dc electric field in the $x$ direction, the EMM along $x$ appears unaffected by the cross terms, while a small amount of EMM along $z$, $\langle E_z^2(\Omega)\rangle < 10^{-7} \langle E_x^2(\Omega)\rangle$, is found for nonzero $\varepsilon_{3,z}$. We anticipate this to be caused by the term proportional to $\varepsilon_{3,z} x^2 z$ in Eq.~\eqref{eq:phi2}, which gives a term proportional $\varepsilon_{3,z} x^2$ in $E_z = -d\phi/dz$ ($x$ has a significant nonzero value due to the dc field in this direction). With a field along $z$, the only observed difference relative to the harmonic potential was a modification of the ratios $\rho_{2,z}$ and $\rho_{3,z}$. 

This motivates us to seek an approximative analytical solution by neglecting the cross terms and considering an electric trap potential proportional to $x_i^2 + \varepsilon_{3,i} x_i^3 + \varepsilon_{4,i} x_i^4$ along the trap axis $i$.  This leads to the nonlinear Mathieu equation
\begin{eqnarray} \label{eq:mathieu-nl}
\frac{d^2 x_i}{d\tau^2} + a_i x_i - 2q_i \cos{2\tau} (x_i + \frac{3}{2}\varepsilon_{3,i} x_i^2 + 2\varepsilon_{4,i} x_i^3) \nonumber \\
= \frac{4 e}{m\Omega^2}  E_{\mathrm{dc},i}.
\end{eqnarray}
We use the \emph{ansatz}
$x_i(\tau)=\sum_{n=-\infty}^\infty b_{2n,i} e^{\mathrm{i} 2n \tau}$,
but nothing in Eq.~\eqref{eq:mathieu-nl} breaks the symmetry between positive and negative $n$, thus $b_{-2n,i}=b_{2n,i}$. Whereas the linear Mathieu equation \eqref{eq:mathieu} led to the recursion relations \eqref{eq:rec}, where the order $n$ coefficient is coupled only to those of order $n+1$ and $n-1$, the nonlinear terms in Eq.~\eqref{eq:mathieu-nl} give rise to coupling to all other coefficients. We solve the equation approximatively up to $n=3$ by using the solution of the linear Mathieu equation \eqref{eq:x-sol} for $x_i$ in the small nonlinear terms. This gives coefficients $b_{2n,i}$ that have fractional corrections of order $\varepsilon_{3,i} c_{0,i}$ and $\varepsilon_{4,i} c_{0,i}^2$ relative to the linear Mathieu coefficients $c_{2n,i}$, with $c_{0,i}$, Eq.~\eqref{eq:c0}, being the mean displacement from the rf null. 

\begin{figure}[t]
\includegraphics[width=1\columnwidth]{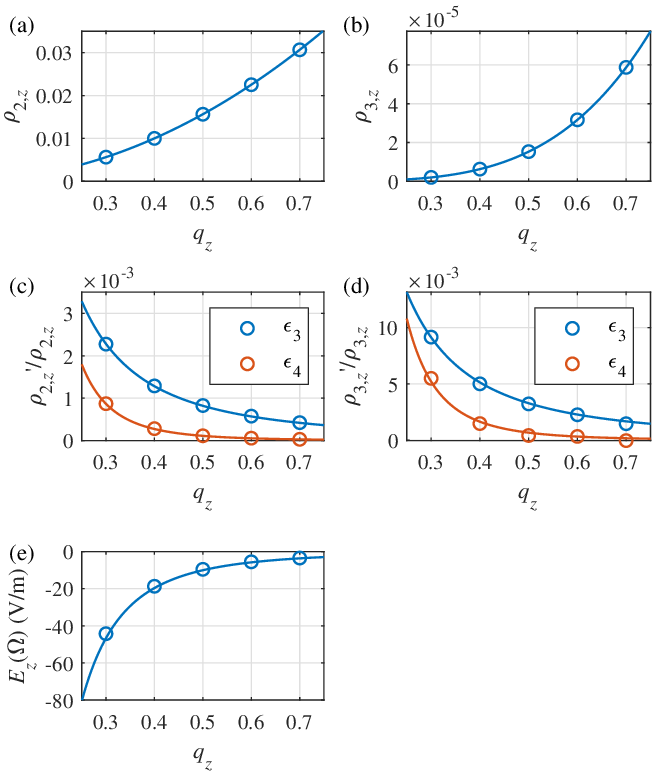}%
\caption{\label{fig:ansiotr}
Comparison of simulated (circles) and analytical (lines) results. (a) $\rho_{2,z}$ and (b) $\rho_{3,z}$ as functions of $q_z$ for a harmonic potential and a \qty{1.6}{kV/m} dc field along $z$. Fractional change in (c) $\rho_{2,z}$ and (d) $\rho_{3,z}$ caused by either $\varepsilon_3 = \qty{2e-5}{\micro\meter^{-1}}$ or $\varepsilon_4 = \qty{2e-7}{\micro\meter^{-2}}$. (e) rf electric field along $z$ caused by a \qty{1.0}{kV/m} dc field along $x$ and $\varepsilon_3 = \qty{2e-5}{\micro\meter^{-1}}$.
}
\end{figure}

In particular, we get the corrections to Eqs.~(2) in the main text as
\begin{subequations} \label{eq:rhop}
\begin{eqnarray} 
  \rho_{2,i}' &=& \rho_{2,i}(1+6 \varepsilon_{3,i} c_{0,i} + 12 \varepsilon_{4,i} c_{0,i}^2), \\
  \rho_{3,i}' &=& \rho_{3,i}(1+24 \varepsilon_{3,i} c_{0,i} + 72 \varepsilon_{4,i} c_{0,i}^2).
\end{eqnarray}
\end{subequations}
These relations were validated by the above mentioned numerical simulations in (py)LIon with $E_{\text{dc},z}=1.6$\;kV/m, see Fig.~\ref{fig:ansiotr}(a--d). The simulation used $\Omega/2\pi = 14.4$\;MHz and $a_i = 0$. Figure~\ref{fig:ansiotr}(e) shows the rf electric field along $z$ caused by a field $E_{\text{dc},x}=1.0$\;kV/m along $x$. Here, the analytical expression is 
\begin{equation}
    E_z(\Omega) = -\frac{m \Omega^2}{8 e} q_z 3 \varepsilon_{3,z} b_{0,x}^2,
\end{equation}
i.e., it is the electric field corresponding to the term proportional to $\varepsilon_{3,z} x^2 z$ in Eq.~\eqref{eq:phi2} with the ion displacement $b_{0,x}$ caused by $E_{\text{dc},x}$ substituted for $x$.

As $\bar{\rho}_3\ll \bar{\rho}_2$, it is for the DSSP uncertainty evaluation sufficient to consider the correction to the latter, for which we get the corrected value
\begin{eqnarray}
\bar{\rho}_2' &\approx& \bar{\rho}_2 + \sum_i u_{\Omega,i}^2 \left(\frac{q_i}{4}\right)^2 \left( 6 \varepsilon_{3,i} c_{0,i} + 12 \varepsilon_{4,i} c_{0,i}^2 \right) \nonumber \\
&\equiv& \bar{\rho}_2 \left(1+ 6 \bar{\varepsilon}_{3} c_{0} + 12 \bar{\varepsilon}_{4} c_{0}^2 \right) \equiv \bar{\rho}_2 + \Delta\bar{\rho}_2,
\end{eqnarray}
where $c_0$ is the total displacement along $\mathbf{u}_\mathrm{d}$ and the second line defines the effective mean coefficients $\bar{\varepsilon}_j$, which are weighted by $u_{\Omega,i}$, $q_i$, and $c_{0,i}$, and $\Delta\bar{\rho}_2$, which is the change in $\bar{\rho}_2$. For the uncertainty estimation, $\Delta\bar{\rho}_2$ is used to write the fraction in Eqs.~(3--4) in the main text as
\begin{equation}
    \frac{1 + \bar{\rho}_2'/4 + \bar{\rho}_3/9}{1 + \bar{\rho}_2' + \bar{\rho}_3} \approx \frac{1 + \bar{\rho}_2/4 + \bar{\rho}_3/9}{1 + \bar{\rho}_2 + \bar{\rho}_3}\left( 1- \frac{3}{4} \Delta\bar{\rho}_2\right).
\end{equation}

The anharmonicity coefficients along the chamber axes $X,Y,Z$ obtained from a finite-element-method (FEM) simulation \cite{sLindvall2022a} are given in Table~\ref{tab:anharm} together with the corresponding effective mean values $\bar{\varepsilon}_j$. Experimentally, we estimate the coefficients by measuring the secular frequencies for different ion displacements along $\mathbf{u}_\mathrm{d}$ and fitting a second-order polynomial to the calculated $q_z$ values. This method assumes radial symmetry, so these values should be taken as a rough estimate. For estimating the DSSP uncertainty, we use worst-case estimates inflated by about a factor of two relative to the larger of the simulated and measured values, see Table~\ref{tab:anharm}.

The magnitude of the anharmonicity corrections to the individual DSSP values in Fig.~4 in the main text, $3/4\, \Delta\bar{\rho}_2 \Delta\alpha_0$, is $(2\ldots5)\times 10^{-45}\;\mathrm{J\,m^2/V^2}$. However, if we instead correct the zero crossing frequencies by $3/8\, \Delta\bar{\rho}_2 \Omega_0$ and repeat the $q_z$ fitting, the change in the obtained DSSP value is only $\num{2e-46}\;\mathrm{J\,m^2/V^2}$. This again demonstrates how the $q_z$ fitting method suppresses nonidealities.

\begin{table}[h]
\centering
  \caption{
    Trap anharmonicity coefficients.
   \label{tab:anharm}}
   \begin{ruledtabular}
  \begin{tabular}{@{} l *2{S[table-format = +2.1e+1]} @{}}
 & {$\quad \varepsilon_{3}$ ($\si{\micro\m}^{-1}$)} & {$\quad \varepsilon_{4}$ ($\si{\micro\m}^{-2}$)}  \\
\midrule
FEM simulation, $X$ & -2.1e-7 & -1.8e-7 \\
FEM simulation, $Y$ & -0.2e-7 & -1.8e-7 \\
FEM simulation, $Z$ & -0.4e-7 &  2.3e-7 \\
FEM, effective mean $\bar{\varepsilon}_j$ & -0.2e-7 &  0.9e-7 \\
Measured  & 8.5e-6 &  0.7e-7 \\
Worst-case estimate  & 2e-5 &  2e-7 \\
  \end{tabular}
  \end{ruledtabular}
\end{table}


\begin{thebibliography}{42}%
\makeatletter
\providecommand \@ifxundefined [1]{%
 \@ifx{#1\undefined}
}%
\providecommand \@ifnum [1]{%
 \ifnum #1\expandafter \@firstoftwo
 \else \expandafter \@secondoftwo
 \fi
}%
\providecommand \@ifx [1]{%
 \ifx #1\expandafter \@firstoftwo
 \else \expandafter \@secondoftwo
 \fi
}%
\providecommand \natexlab [1]{#1}%
\providecommand \enquote  [1]{``#1''}%
\providecommand \bibnamefont  [1]{#1}%
\providecommand \bibfnamefont [1]{#1}%
\providecommand \citenamefont [1]{#1}%
\providecommand \href@noop [0]{\@secondoftwo}%
\providecommand \href [0]{\begingroup \@sanitize@url \@href}%
\providecommand \@href[1]{\@@startlink{#1}\@@href}%
\providecommand \@@href[1]{\endgroup#1\@@endlink}%
\providecommand \@sanitize@url [0]{\catcode `\\12\catcode `\$12\catcode `\&12\catcode `\#12\catcode `\^12\catcode `\_12\catcode `\%12\relax}%
\providecommand \@@startlink[1]{}%
\providecommand \@@endlink[0]{}%
\providecommand \url  [0]{\begingroup\@sanitize@url \@url }%
\providecommand \@url [1]{\endgroup\@href {#1}{\urlprefix }}%
\providecommand \urlprefix  [0]{URL }%
\providecommand \Eprint [0]{\href }%
\providecommand \doibase [0]{https://doi.org/}%
\providecommand \selectlanguage [0]{\@gobble}%
\providecommand \bibinfo  [0]{\@secondoftwo}%
\providecommand \bibfield  [0]{\@secondoftwo}%
\providecommand \translation [1]{[#1]}%
\providecommand \BibitemOpen [0]{}%
\providecommand \bibitemStop [0]{}%
\providecommand \bibitemNoStop [0]{.\EOS\space}%
\providecommand \EOS [0]{\spacefactor3000\relax}%
\providecommand \BibitemShut  [1]{\csname bibitem#1\endcsname}%
\let\auto@bib@innerbib\@empty
\bibitem [{\citenamefont {Ludlow}\ \emph {et~al.}(2015)\citenamefont {Ludlow}, \citenamefont {Boyd}, \citenamefont {Ye}, \citenamefont {Peik},\ and\ \citenamefont {Schmidt}}]{Ludlow2015b}%
  \BibitemOpen
  \bibfield  {author} {\bibinfo {author} {\bibfnamefont {A.~D.}\ \bibnamefont {Ludlow}}, \bibinfo {author} {\bibfnamefont {M.~M.}\ \bibnamefont {Boyd}}, \bibinfo {author} {\bibfnamefont {J.}~\bibnamefont {Ye}}, \bibinfo {author} {\bibfnamefont {E.}~\bibnamefont {Peik}},\ and\ \bibinfo {author} {\bibfnamefont {P.~O.}\ \bibnamefont {Schmidt}},\ }\bibfield  {title} {\bibinfo {title} {{O}ptical atomic clocks},\ }\href {https://doi.org/10.1103/RevModPhys.87.637} {\bibfield  {journal} {\bibinfo  {journal} {Rev. Mod. Phys.}\ }\textbf {\bibinfo {volume} {87}},\ \bibinfo {pages} {637} (\bibinfo {year} {2015})}\BibitemShut {NoStop}%
\bibitem [{\citenamefont {Safronova}\ \emph {et~al.}(2018)\citenamefont {Safronova}, \citenamefont {Budker}, \citenamefont {DeMille}, \citenamefont {Kimball}, \citenamefont {Derevianko},\ and\ \citenamefont {Clark}}]{Safronova2018b}%
  \BibitemOpen
  \bibfield  {author} {\bibinfo {author} {\bibfnamefont {M.~S.}\ \bibnamefont {Safronova}}, \bibinfo {author} {\bibfnamefont {D.}~\bibnamefont {Budker}}, \bibinfo {author} {\bibfnamefont {D.}~\bibnamefont {DeMille}}, \bibinfo {author} {\bibfnamefont {D.~F.~J.}\ \bibnamefont {Kimball}}, \bibinfo {author} {\bibfnamefont {A.}~\bibnamefont {Derevianko}},\ and\ \bibinfo {author} {\bibfnamefont {C.~W.}\ \bibnamefont {Clark}},\ }\bibfield  {title} {\bibinfo {title} {Search for new physics with atoms and molecules},\ }\href {https://doi.org/10.1103/RevModPhys.90.025008} {\bibfield  {journal} {\bibinfo  {journal} {Rev. Mod. Phys.}\ }\textbf {\bibinfo {volume} {90}},\ \bibinfo {pages} {025008} (\bibinfo {year} {2018})}\BibitemShut {NoStop}%
\bibitem [{\citenamefont {Mehlst\"aubler}\ \emph {et~al.}(2018)\citenamefont {Mehlst\"aubler}, \citenamefont {Grosche}, \citenamefont {Lisdat}, \citenamefont {Schmidt},\ and\ \citenamefont {Denker}}]{Mehlstaubler2018a}%
  \BibitemOpen
  \bibfield  {author} {\bibinfo {author} {\bibfnamefont {T.~E.}\ \bibnamefont {Mehlst\"aubler}}, \bibinfo {author} {\bibfnamefont {G.}~\bibnamefont {Grosche}}, \bibinfo {author} {\bibfnamefont {C.}~\bibnamefont {Lisdat}}, \bibinfo {author} {\bibfnamefont {P.~O.}\ \bibnamefont {Schmidt}},\ and\ \bibinfo {author} {\bibfnamefont {H.}~\bibnamefont {Denker}},\ }\bibfield  {title} {\bibinfo {title} {Atomic clocks for geodesy},\ }\href {https://doi.org/10.1088/1361-6633/aab409} {\bibfield  {journal} {\bibinfo  {journal} {Rep. Prog. Phys.}\ }\textbf {\bibinfo {volume} {81}},\ \bibinfo {pages} {064401} (\bibinfo {year} {2018})}\BibitemShut {NoStop}%
\bibitem [{\citenamefont {Dimarcq}\ \emph {et~al.}(2024)\citenamefont {Dimarcq}, \citenamefont {Gertsvolf}, \citenamefont {Mileti}, \citenamefont {Bize}, \citenamefont {Oates}, \citenamefont {Peik}, \citenamefont {Calonico}, \citenamefont {Ido}, \citenamefont {Tavella}, \citenamefont {Meynadier}, \citenamefont {Petit}, \citenamefont {Panfilo}, \citenamefont {Bartholomew}, \citenamefont {Defraigne}, \citenamefont {Donley}, \citenamefont {Hedekvist}, \citenamefont {Sesia}, \citenamefont {Wouters}, \citenamefont {Dube}, \citenamefont {Fang}, \citenamefont {Levi}, \citenamefont {Lodewyck}, \citenamefont {Margolis}, \citenamefont {Newell}, \citenamefont {Slyusarev}, \citenamefont {Weyers}, \citenamefont {Uzan}, \citenamefont {Yasuda}, \citenamefont {Yu}, \citenamefont {Rieck}, \citenamefont {Schnatz}, \citenamefont {Hanado}, \citenamefont {Fujieda}, \citenamefont {Pottie}, \citenamefont {Hanssen}, \citenamefont {Malimon},\ and\ \citenamefont {Ashby}}]{Dimarcq2024a}%
  \BibitemOpen
  \bibfield  {author} {\bibinfo {author} {\bibfnamefont {N.}~\bibnamefont {Dimarcq} \emph{et~al.}},\ }\bibfield  {title} {\bibinfo {title} {Roadmap towards the redefinition of the second},\ }\href {https://doi.org/10.1088/1681-7575/ad17d2} {\bibfield  {journal} {\bibinfo  {journal} {Metrologia}\ }\textbf {\bibinfo {volume} {61}},\ \bibinfo {pages} {012001} (\bibinfo {year} {2024})} \BibitemShut {NoStop}%
\bibitem [{\citenamefont {{Safronova}}\ \emph {et~al.}(2012)\citenamefont {{Safronova}}, \citenamefont {{Kozlov}},\ and\ \citenamefont {{Clark}}}]{Safronova2012b}%
  \BibitemOpen
  \bibfield  {author} {\bibinfo {author} {\bibfnamefont {M.~S.}\ \bibnamefont {{Safronova}}}, \bibinfo {author} {\bibfnamefont {M.~G.}\ \bibnamefont {{Kozlov}}},\ and\ \bibinfo {author} {\bibfnamefont {C.~W.}\ \bibnamefont {{Clark}}},\ }\bibfield  {title} {\bibinfo {title} {Blackbody radiation shifts in optical atomic clocks},\ }\href {https://doi.org/10.1109/TUFFC.2012.2213} {\bibfield  {journal} {\bibinfo  {journal} {IEEE Trans. Ultrason., Ferroelectr., Freq. Control}\ }\textbf {\bibinfo {volume} {59}},\ \bibinfo {pages} {439} (\bibinfo {year} {2012})}\BibitemShut {NoStop}%
\bibitem [{\citenamefont {Huang}\ \emph {et~al.}(2022)\citenamefont {Huang}, \citenamefont {Zhang}, \citenamefont {Zeng}, \citenamefont {Hao}, \citenamefont {Ma}, \citenamefont {Zhang}, \citenamefont {Guan}, \citenamefont {Chen}, \citenamefont {Wang},\ and\ \citenamefont {Gao}}]{Huang2022a}%
  \BibitemOpen
  \bibfield  {author} {\bibinfo {author} {\bibfnamefont {Y.}~\bibnamefont {Huang}}, \bibinfo {author} {\bibfnamefont {B.}~\bibnamefont {Zhang}}, \bibinfo {author} {\bibfnamefont {M.}~\bibnamefont {Zeng}}, \bibinfo {author} {\bibfnamefont {Y.}~\bibnamefont {Hao}}, \bibinfo {author} {\bibfnamefont {Z.}~\bibnamefont {Ma}}, \bibinfo {author} {\bibfnamefont {H.}~\bibnamefont {Zhang}}, \bibinfo {author} {\bibfnamefont {H.}~\bibnamefont {Guan}}, \bibinfo {author} {\bibfnamefont {Z.}~\bibnamefont {Chen}}, \bibinfo {author} {\bibfnamefont {M.}~\bibnamefont {Wang}},\ and\ \bibinfo {author} {\bibfnamefont {K.}~\bibnamefont {Gao}},\ }\bibfield  {title} {\bibinfo {title} {Liquid-nitrogen-cooled {Ca}$^ +$ optical clock with systematic uncertainty of $3 \times 10^{-18}$},\ }\href {https://doi.org/10.1103/physrevapplied.17.034041} {\bibfield  {journal} {\bibinfo  {journal} {Phys. Rev. Appl.}\ }\textbf {\bibinfo {volume} {17}},\ \bibinfo {pages} {034041} (\bibinfo {year} {2022})} \BibitemShut {NoStop}%
\bibitem [{\citenamefont {Huntemann}\ \emph {et~al.}(2016)\citenamefont {Huntemann}, \citenamefont {Sanner}, \citenamefont {Lipphardt}, \citenamefont {Tamm},\ and\ \citenamefont {Peik}}]{Huntemann2016a}%
  \BibitemOpen
  \bibfield  {author} {\bibinfo {author} {\bibfnamefont {N.}~\bibnamefont {Huntemann}}, \bibinfo {author} {\bibfnamefont {C.}~\bibnamefont {Sanner}}, \bibinfo {author} {\bibfnamefont {B.}~\bibnamefont {Lipphardt}}, \bibinfo {author} {\bibfnamefont {C.}~\bibnamefont {Tamm}},\ and\ \bibinfo {author} {\bibfnamefont {E.}~\bibnamefont {Peik}},\ }\bibfield  {title} {\bibinfo {title} {{S}ingle-{I}on {A}tomic {C}lock with $3\times 10^{-18}$ {S}ystematic {U}ncertainty},\ }\href {https://doi.org/10.1103/PhysRevLett.116.063001} {\bibfield  {journal} {\bibinfo  {journal} {Phys. Rev. Lett.}\ }\textbf {\bibinfo {volume} {116}},\ \bibinfo {pages} {063001} (\bibinfo {year} {2016})}\BibitemShut {NoStop}%
\bibitem [{\citenamefont {Baynham}\ \emph {et~al.}(2018)\citenamefont {Baynham}, \citenamefont {Curtis}, \citenamefont {Godun}, \citenamefont {Jones}, \citenamefont {Nisbet-Jones}, \citenamefont {Baird}, \citenamefont {Bongs}, \citenamefont {Gill}, \citenamefont {Fordell}, \citenamefont {Hieta}, \citenamefont {Lindvall}, \citenamefont {Spidell},\ and\ \citenamefont {Lehman}}]{Baynham2018b}%
  \BibitemOpen
  \bibfield  {author} {\bibinfo {author} {\bibfnamefont {C.~F.~A.}\ \bibnamefont {Baynham}}, \bibinfo {author} {\bibfnamefont {E.~A.}\ \bibnamefont {Curtis}}, \bibinfo {author} {\bibfnamefont {R.~M.}\ \bibnamefont {Godun}}, \bibinfo {author} {\bibfnamefont {J.~M.}\ \bibnamefont {Jones}}, \bibinfo {author} {\bibfnamefont {P.~B.~R.}\ \bibnamefont {Nisbet-Jones}}, \bibinfo {author} {\bibfnamefont {P.~E.~G.}\ \bibnamefont {Baird}}, \bibinfo {author} {\bibfnamefont {K.}~\bibnamefont {Bongs}}, \bibinfo {author} {\bibfnamefont {P.}~\bibnamefont {Gill}}, \bibinfo {author} {\bibfnamefont {T.}~\bibnamefont {Fordell}}, \bibinfo {author} {\bibfnamefont {T.}~\bibnamefont {Hieta}}, \bibinfo {author} {\bibfnamefont {T.}~\bibnamefont {Lindvall}}, \bibinfo {author} {\bibfnamefont {M.~T.}\ \bibnamefont {Spidell}},\ and\ \bibinfo {author} {\bibfnamefont {J.~H.}\ \bibnamefont {Lehman}},\ }\bibfield  {title} {\bibinfo {title} {Measurement of differential polarizabilities at a mid-infrared wavelength in $^{171}\mathrm{Yb}^+$}\
  }\href {https://doi.org/10.48550/ARXIV.1801.10134} {10.48550/ARXIV.1801.10134} (\bibinfo {year} {2018}),\ \Eprint {https://arxiv.org/abs/1801.10134} {arXiv:1801.10134 [physics.atom-ph]} \BibitemShut {NoStop}%
\bibitem [{\citenamefont {Arnold}\ \emph {et~al.}(2018)\citenamefont {Arnold}, \citenamefont {Kaewuam}, \citenamefont {Roy}, \citenamefont {Tan},\ and\ \citenamefont {Barrett}}]{Arnold2018a}%
  \BibitemOpen
  \bibfield  {author} {\bibinfo {author} {\bibfnamefont {K.~J.}\ \bibnamefont {Arnold}}, \bibinfo {author} {\bibfnamefont {R.}~\bibnamefont {Kaewuam}}, \bibinfo {author} {\bibfnamefont {A.}~\bibnamefont {Roy}}, \bibinfo {author} {\bibfnamefont {T.~R.}\ \bibnamefont {Tan}},\ and\ \bibinfo {author} {\bibfnamefont {M.~D.}\ \bibnamefont {Barrett}},\ }\bibfield  {title} {\bibinfo {title} {Blackbody radiation shift assessment for a lutetium ion clock},\ }\href {https://doi.org/10.1038/s41467-018-04079-x} {\bibfield  {journal} {\bibinfo  {journal} {Nat. Commun.}\ }\textbf {\bibinfo {volume} {9}},\ \bibinfo {pages} {1650} (\bibinfo {year} {2018})}\BibitemShut {NoStop}%
\bibitem [{\citenamefont {Arnold}\ \emph {et~al.}(2019)\citenamefont {Arnold}, \citenamefont {Kaewuam}, \citenamefont {Tan}, \citenamefont {Porsev}, \citenamefont {Safronova},\ and\ \citenamefont {Barrett}}]{Arnold2019a}%
  \BibitemOpen
  \bibfield  {author} {\bibinfo {author} {\bibfnamefont {K.~J.}\ \bibnamefont {Arnold}}, \bibinfo {author} {\bibfnamefont {R.}~\bibnamefont {Kaewuam}}, \bibinfo {author} {\bibfnamefont {T.~R.}\ \bibnamefont {Tan}}, \bibinfo {author} {\bibfnamefont {S.~G.}\ \bibnamefont {Porsev}}, \bibinfo {author} {\bibfnamefont {M.~S.}\ \bibnamefont {Safronova}},\ and\ \bibinfo {author} {\bibfnamefont {M.~D.}\ \bibnamefont {Barrett}},\ }\bibfield  {title} {\bibinfo {title} {Dynamic polarizability measurements with $^{176}\mathrm{Lu}^+$},\ }\href {https://doi.org/10.1103/PhysRevA.99.012510} {\bibfield  {journal} {\bibinfo  {journal} {Phys. Rev. A}\ }\textbf {\bibinfo {volume} {99}},\ \bibinfo {pages} {012510} (\bibinfo {year} {2019})}\BibitemShut {NoStop}%
\bibitem [{\citenamefont {Brewer}\ \emph {et~al.}(2019)\citenamefont {Brewer}, \citenamefont {Chen}, \citenamefont {Hankin}, \citenamefont {Clements}, \citenamefont {Chou}, \citenamefont {Wineland}, \citenamefont {Hume},\ and\ \citenamefont {Leibrandt}}]{Brewer2019a}%
  \BibitemOpen
  \bibfield  {author} {\bibinfo {author} {\bibfnamefont {S.~M.}\ \bibnamefont {Brewer}}, \bibinfo {author} {\bibfnamefont {J.-S.}\ \bibnamefont {Chen}}, \bibinfo {author} {\bibfnamefont {A.~M.}\ \bibnamefont {Hankin}}, \bibinfo {author} {\bibfnamefont {E.~R.}\ \bibnamefont {Clements}}, \bibinfo {author} {\bibfnamefont {C.~W.}\ \bibnamefont {Chou}}, \bibinfo {author} {\bibfnamefont {D.~J.}\ \bibnamefont {Wineland}}, \bibinfo {author} {\bibfnamefont {D.~B.}\ \bibnamefont {Hume}},\ and\ \bibinfo {author} {\bibfnamefont {D.~R.}\ \bibnamefont {Leibrandt}},\ }\bibfield  {title} {\bibinfo {title} {$^{27}${Al}$^{+}$ quantum-logic clock with a systematic uncertainty below ${10}^{-18}$},\ }\href {https://doi.org/10.1103/PhysRevLett.123.033201} {\bibfield  {journal} {\bibinfo  {journal} {Phys. Rev. Lett.}\ }\textbf {\bibinfo {volume} {123}},\ \bibinfo {pages} {033201} (\bibinfo {year} {2019})}\BibitemShut {NoStop}%
\bibitem [{\citenamefont {Barrett}\ \emph {et~al.}(2019)\citenamefont {Barrett}, \citenamefont {Arnold},\ and\ \citenamefont {Safronova}}]{Barrett2019a}%
  \BibitemOpen
  \bibfield  {author} {\bibinfo {author} {\bibfnamefont {M.~D.}\ \bibnamefont {Barrett}}, \bibinfo {author} {\bibfnamefont {K.~J.}\ \bibnamefont {Arnold}},\ and\ \bibinfo {author} {\bibfnamefont {M.~S.}\ \bibnamefont {Safronova}},\ }\bibfield  {title} {\bibinfo {title} {Polarizability assessments of ion-based optical clocks},\ }\href {https://doi.org/10.1103/PhysRevA.100.043418} {\bibfield  {journal} {\bibinfo  {journal} {Phys. Rev. A}\ }\textbf {\bibinfo {volume} {100}},\ \bibinfo {pages} {043418} (\bibinfo {year} {2019})}\BibitemShut {NoStop}%
\bibitem [{\citenamefont {Huang}\ \emph {et~al.}(2024)\citenamefont {Huang}, \citenamefont {Wang}, \citenamefont {Chen}, \citenamefont {Li}, \citenamefont {Zhang}, \citenamefont {Zhang}, \citenamefont {Tang}, \citenamefont {Shi}, \citenamefont {Guan},\ and\ \citenamefont {Gao}}]{Huang2024a}%
  \BibitemOpen
  \bibfield  {author} {\bibinfo {author} {\bibfnamefont {Y.}~\bibnamefont {Huang}}, \bibinfo {author} {\bibfnamefont {M.}~\bibnamefont {Wang}}, \bibinfo {author} {\bibfnamefont {Z.}~\bibnamefont {Chen}}, \bibinfo {author} {\bibfnamefont {C.}~\bibnamefont {Li}}, \bibinfo {author} {\bibfnamefont {H.}~\bibnamefont {Zhang}}, \bibinfo {author} {\bibfnamefont {B.}~\bibnamefont {Zhang}}, \bibinfo {author} {\bibfnamefont {L.}~\bibnamefont {Tang}}, \bibinfo {author} {\bibfnamefont {T.}~\bibnamefont {Shi}}, \bibinfo {author} {\bibfnamefont {H.}~\bibnamefont {Guan}},\ and\ \bibinfo {author} {\bibfnamefont {K.-L.}\ \bibnamefont {Gao}},\ }\bibfield  {title} {\bibinfo {title} {Measurement of infrared magic wavelength for an all-optical trapping of $^{40}${Ca}$^+$ ion clock},\ }\href {https://doi.org/10.1088/1367-2630/ad3ea8} {\bibfield  {journal} {\bibinfo  {journal} {New J. Phys.}\ }\textbf {\bibinfo {volume} {26}},\ \bibinfo {pages} {043021} (\bibinfo {year} {2024})}\BibitemShut {NoStop}%
\bibitem [{\citenamefont {Barrett}\ and\ \citenamefont {Arnold}(2025)}]{Barrett2025a}%
  \BibitemOpen
  \bibfield  {author} {\bibinfo {author} {\bibfnamefont {M.~D.}\ \bibnamefont {Barrett}}\ and\ \bibinfo {author} {\bibfnamefont {K.~J.}\ \bibnamefont {Arnold}},\ }\bibfield  {title} {\bibinfo {title} {An extrapolation method for polarisability assessments of ion-based optical clocks},\ }\href {https://doi.org/10.1088/1367-2630/ada4d1} {\bibfield  {journal} {\bibinfo  {journal} {New J. Phys.}\ }\textbf {\bibinfo {volume} {27}},\ \bibinfo {pages} {013005} (\bibinfo {year} {2025})} \BibitemShut {NoStop}%
\bibitem [{\citenamefont {Dub\'e}\ \emph {et~al.}(2014)\citenamefont {Dub\'e}, \citenamefont {Madej}, \citenamefont {Tibbo},\ and\ \citenamefont {Bernard}}]{Dube2014a}%
  \BibitemOpen
  \bibfield  {author} {\bibinfo {author} {\bibfnamefont {P.}~\bibnamefont {Dub\'e}}, \bibinfo {author} {\bibfnamefont {A.~A.}\ \bibnamefont {Madej}}, \bibinfo {author} {\bibfnamefont {M.}~\bibnamefont {Tibbo}},\ and\ \bibinfo {author} {\bibfnamefont {J.~E.}\ \bibnamefont {Bernard}},\ }\bibfield  {title} {\bibinfo {title} {{H}igh-{A}ccuracy {M}easurement of the {D}ifferential {S}calar {P}olarizability of a $^{88}${Sr}$^{+}$ {C}lock {U}sing the {T}ime-{D}ilation {E}ffect},\ }\href {https://doi.org/10.1103/PhysRevLett.112.173002} {\bibfield  {journal} {\bibinfo  {journal} {Phys. Rev. Lett.}\ }\textbf {\bibinfo {volume} {112}},\ \bibinfo {pages} {173002} (\bibinfo {year} {2014})}\BibitemShut {NoStop}%
\bibitem [{\citenamefont {Huang}\ \emph {et~al.}(2019)\citenamefont {Huang}, \citenamefont {Guan}, \citenamefont {Zeng}, \citenamefont {Tang},\ and\ \citenamefont {Gao}}]{Huang2019a}%
  \BibitemOpen
  \bibfield  {author} {\bibinfo {author} {\bibfnamefont {Y.}~\bibnamefont {Huang}}, \bibinfo {author} {\bibfnamefont {H.}~\bibnamefont {Guan}}, \bibinfo {author} {\bibfnamefont {M.}~\bibnamefont {Zeng}}, \bibinfo {author} {\bibfnamefont {L.}~\bibnamefont {Tang}},\ and\ \bibinfo {author} {\bibfnamefont {K.}~\bibnamefont {Gao}},\ }\bibfield  {title} {\bibinfo {title} {$^{40}${Ca}$^{+}$ ion optical clock with micromotion-induced shifts below $1\times 10^{-18}$},\ }\href {https://doi.org/10.1103/PhysRevA.99.011401} {\bibfield  {journal} {\bibinfo  {journal} {Phys. Rev. A}\ }\textbf {\bibinfo {volume} {99}},\ \bibinfo {pages} {011401} (\bibinfo {year} {2019})}\BibitemShut {NoStop}%
\bibitem [{\citenamefont {Wolf}(2024)}]{Wolf2024a}%
  \BibitemOpen
  \bibfield  {author} {\bibinfo {author} {\bibfnamefont {F.}~\bibnamefont {Wolf}},\ }\bibfield  {title} {\bibinfo {title} {Scheme for quantum-logic based transfer of accuracy in polarizability measurement for trapped ions using a moving optical lattice},\ }\href {https://doi.org/10.1103/physrevlett.132.083202} {\bibfield  {journal} {\bibinfo  {journal} {Phys. Rev. Lett.}\ }\textbf {\bibinfo {volume} {132}},\ \bibinfo {pages} {083202} (\bibinfo {year} {2024})} \BibitemShut {NoStop}%
\bibitem [{\citenamefont {Wei}\ \emph {et~al.}(2024)\citenamefont {Wei}, \citenamefont {Chao}, \citenamefont {Cui}, \citenamefont {Li}, \citenamefont {Yu}, \citenamefont {Zhang}, \citenamefont {Shu}, \citenamefont {Cao},\ and\ \citenamefont {Huang}}]{Wei2024a}%
  \BibitemOpen
  \bibfield  {author} {\bibinfo {author} {\bibfnamefont {Y.-F.}\ \bibnamefont {Wei}}, \bibinfo {author} {\bibfnamefont {S.-J.}\ \bibnamefont {Chao}}, \bibinfo {author} {\bibfnamefont {K.-F.}\ \bibnamefont {Cui}}, \bibinfo {author} {\bibfnamefont {C.-B.}\ \bibnamefont {Li}}, \bibinfo {author} {\bibfnamefont {S.-C.}\ \bibnamefont {Yu}}, \bibinfo {author} {\bibfnamefont {H.}~\bibnamefont {Zhang}}, \bibinfo {author} {\bibfnamefont {H.-L.}\ \bibnamefont {Shu}}, \bibinfo {author} {\bibfnamefont {J.}~\bibnamefont {Cao}},\ and\ \bibinfo {author} {\bibfnamefont {X.-R.}\ \bibnamefont {Huang}},\ }\bibfield  {title} {\bibinfo {title} {Improved measurement of the differential polarizability using co-trapped ions},\ }\href {https://doi.org/10.1103/physrevlett.133.033001} {\bibfield  {journal} {\bibinfo  {journal} {Phys. Rev. Lett.}\ }\textbf {\bibinfo {volume} {133}},\ \bibinfo {pages} {033001} (\bibinfo {year} {2024})}\BibitemShut {NoStop}%
\bibitem [{\citenamefont {Akerman}\ and\ \citenamefont {Ozeri}(2025)}]{Akerman2025a}%
  \BibitemOpen
  \bibfield  {author} {\bibinfo {author} {\bibfnamefont {N.}~\bibnamefont {Akerman}}\ and\ \bibinfo {author} {\bibfnamefont {R.}~\bibnamefont {Ozeri}},\ }\bibfield  {title} {\bibinfo {title} {Operating a multi-ion clock with dynamical decoupling},\ }\href {https://doi.org/10.1103/physrevlett.134.013201} {\bibfield  {journal} {\bibinfo  {journal} {Phys. Rev. Lett.}\ }\textbf {\bibinfo {volume} {134}},\ \bibinfo {pages} {013201} (\bibinfo {year} {2025})} \BibitemShut {NoStop}%
\bibitem [{\citenamefont {Loh}\ \emph {et~al.}(2025)\citenamefont {Loh}, \citenamefont {Reens}, \citenamefont {Kharas}, \citenamefont {Sumant}, \citenamefont {Belanger}, \citenamefont {Maxson}, \citenamefont {Medeiros}, \citenamefont {Setzer}, \citenamefont {Gray}, \citenamefont {DeBry}, \citenamefont {Bruzewicz}, \citenamefont {Plant}, \citenamefont {Liddell}, \citenamefont {West}, \citenamefont {Doshi}, \citenamefont {Roychowdhury}, \citenamefont {Kim}, \citenamefont {Braje}, \citenamefont {Juodawlkis}, \citenamefont {Chiaverini},\ and\ \citenamefont {McConnell}}]{Loh2025a}%
  \BibitemOpen
  \bibfield  {author} {\bibinfo {author} {\bibfnamefont {W.}~\bibnamefont {Loh} \emph{et~al.}},\ }\bibfield  {title} {\bibinfo {title} {Optical atomic clock interrogation using an integrated spiral cavity laser},\ }\href {https://doi.org/10.1038/s41566-024-01588-8} {\bibfield  {journal} {\bibinfo  {journal} {Nat. Photonics}\ }\textbf {\bibinfo {volume} {19}},\ \bibinfo {pages} {277} (\bibinfo {year} {2025})} \BibitemShut {NoStop}%
\bibitem [{\citenamefont {Spampinato}\ \emph {et~al.}(2024)\citenamefont {Spampinato}, \citenamefont {Stacey}, \citenamefont {Mulholland}, \citenamefont {Robertson}, \citenamefont {Klein}, \citenamefont {Huang}, \citenamefont {Barwood},\ and\ \citenamefont {Gill}}]{Spampinato2024a}%
  \BibitemOpen
  \bibfield  {author} {\bibinfo {author} {\bibfnamefont {A.}~\bibnamefont {Spampinato}}, \bibinfo {author} {\bibfnamefont {J.}~\bibnamefont {Stacey}}, \bibinfo {author} {\bibfnamefont {S.}~\bibnamefont {Mulholland}}, \bibinfo {author} {\bibfnamefont {B.~I.}\ \bibnamefont {Robertson}}, \bibinfo {author} {\bibfnamefont {H.~A.}\ \bibnamefont {Klein}}, \bibinfo {author} {\bibfnamefont {G.}~\bibnamefont {Huang}}, \bibinfo {author} {\bibfnamefont {G.~P.}\ \bibnamefont {Barwood}},\ and\ \bibinfo {author} {\bibfnamefont {P.}~\bibnamefont {Gill}},\ }\bibfield  {title} {\bibinfo {title} {An ion trap design for a space-deployable strontium-ion optical clock},\ }\href {https://doi.org/10.1098/rspa.2023.0593} {\bibfield  {journal} {\bibinfo  {journal} {Proc. R. Soc. A}\ }\textbf {\bibinfo {volume} {480}},\ \bibinfo {pages} {20230593} (\bibinfo {year} {2024})}\BibitemShut {NoStop}%
\bibitem [{\citenamefont {Steinel}\ \emph {et~al.}(2023)\citenamefont {Steinel}, \citenamefont {Shao}, \citenamefont {Filzinger}, \citenamefont {Lipphardt}, \citenamefont {Brinkmann}, \citenamefont {Didier}, \citenamefont {Mehlstäubler}, \citenamefont {Lindvall}, \citenamefont {Peik},\ and\ \citenamefont {Huntemann}}]{Steinel2023a}%
  \BibitemOpen
  \bibfield  {author} {\bibinfo {author} {\bibfnamefont {M.}~\bibnamefont {Steinel}}, \bibinfo {author} {\bibfnamefont {H.}~\bibnamefont {Shao}}, \bibinfo {author} {\bibfnamefont {M.}~\bibnamefont {Filzinger}}, \bibinfo {author} {\bibfnamefont {B.}~\bibnamefont {Lipphardt}}, \bibinfo {author} {\bibfnamefont {M.}~\bibnamefont {Brinkmann}}, \bibinfo {author} {\bibfnamefont {A.}~\bibnamefont {Didier}}, \bibinfo {author} {\bibfnamefont {T.~E.}\ \bibnamefont {Mehlstäubler}}, \bibinfo {author} {\bibfnamefont {T.}~\bibnamefont {Lindvall}}, \bibinfo {author} {\bibfnamefont {E.}~\bibnamefont {Peik}},\ and\ \bibinfo {author} {\bibfnamefont {N.}~\bibnamefont {Huntemann}},\ }\bibfield  {title} {\bibinfo {title} {Evaluation of a $^{88}${Sr}$^+$ optical clock with a direct measurement of the blackbody radiation shift and determination of the clock frequency},\ }\href {https://doi.org/10.1103/physrevlett.131.083002} {\bibfield  {journal} {\bibinfo  {journal} {Phys. Rev. Lett.}\ }\textbf {\bibinfo {volume} {131}},\ \bibinfo
  {pages} {083002} (\bibinfo {year} {2023})} \BibitemShut {NoStop}%
\bibitem [{\citenamefont {Dub\'e}\ \emph {et~al.}(2021)\citenamefont {Dub\'e}, \citenamefont {Kato}, \citenamefont {Bernard},\ and\ \citenamefont {Jian}}]{Dube2021a}%
  \BibitemOpen
  \bibfield  {author} {\bibinfo {author} {\bibfnamefont {P.}~\bibnamefont {Dub\'e}}, \bibinfo {author} {\bibfnamefont {K.}~\bibnamefont {Kato}}, \bibinfo {author} {\bibfnamefont {J.}~\bibnamefont {Bernard}},\ and\ \bibinfo {author} {\bibfnamefont {B.}~\bibnamefont {Jian}},\ }\bibfield  {title} {\bibinfo {title} {Progress towards a transportable and high-accuracy {Sr}$^+$ ion clock at {NRC}},\ }in\ \href {https://doi.org/10.1109/eftf/ifcs52194.2021.9604288} {\emph {\bibinfo {booktitle} {2021 Joint Conference of the European Frequency and Time Forum and {IEEE} International Frequency Control Symposium ({EFTF}/{IFCS})}}}\ (\bibinfo  {publisher} {{IEEE}},\ \bibinfo {year} {2021})\BibitemShut {NoStop}%
\bibitem [{\citenamefont {Arnold}\ \emph {et~al.}(2020)\citenamefont {Arnold}, \citenamefont {Kaewuam}, \citenamefont {Chanu}, \citenamefont {Tan}, \citenamefont {Zhang},\ and\ \citenamefont {Barrett}}]{Arnold2020a}%
  \BibitemOpen
  \bibfield  {author} {\bibinfo {author} {\bibfnamefont {K.~J.}\ \bibnamefont {Arnold}}, \bibinfo {author} {\bibfnamefont {R.}~\bibnamefont {Kaewuam}}, \bibinfo {author} {\bibfnamefont {S.~R.}\ \bibnamefont {Chanu}}, \bibinfo {author} {\bibfnamefont {T.~R.}\ \bibnamefont {Tan}}, \bibinfo {author} {\bibfnamefont {Z.}~\bibnamefont {Zhang}},\ and\ \bibinfo {author} {\bibfnamefont {M.~D.}\ \bibnamefont {Barrett}},\ }\bibfield  {title} {\bibinfo {title} {Precision measurements of the $^{138}${Ba}$^{+}$ $6s{^{2}S}_{1/2}\ensuremath{-}5d{^{2}D}_{5/2}$ clock transition},\ }\href {https://doi.org/10.1103/PhysRevLett.124.193001} {\bibfield  {journal} {\bibinfo  {journal} {Phys. Rev. Lett.}\ }\textbf {\bibinfo {volume} {124}},\ \bibinfo {pages} {193001} (\bibinfo {year} {2020})}\BibitemShut {NoStop}%
\bibitem [{\citenamefont {Holliman}\ \emph {et~al.}(2022)\citenamefont {Holliman}, \citenamefont {Fan}, \citenamefont {Contractor}, \citenamefont {Brewer},\ and\ \citenamefont {Jayich}}]{Holliman2022a}%
  \BibitemOpen
  \bibfield  {author} {\bibinfo {author} {\bibfnamefont {C.~A.}\ \bibnamefont {Holliman}}, \bibinfo {author} {\bibfnamefont {M.}~\bibnamefont {Fan}}, \bibinfo {author} {\bibfnamefont {A.}~\bibnamefont {Contractor}}, \bibinfo {author} {\bibfnamefont {S.~M.}\ \bibnamefont {Brewer}},\ and\ \bibinfo {author} {\bibfnamefont {A.~M.}\ \bibnamefont {Jayich}},\ }\bibfield  {title} {\bibinfo {title} {Radium ion optical clock},\ }\href {https://doi.org/10.1103/physrevlett.128.033202} {\bibfield  {journal} {\bibinfo  {journal} {Phys. Rev. Lett.}\ }\textbf {\bibinfo {volume} {128}},\ \bibinfo {pages} {033202} (\bibinfo {year} {2022})} \BibitemShut {NoStop}%
\bibitem [{\citenamefont {Arnold}\ \emph {et~al.}(2024)\citenamefont {Arnold}, \citenamefont {Bustabad}, \citenamefont {Qi}, \citenamefont {Qichen}, \citenamefont {Zhang}, \citenamefont {Zhao},\ and\ \citenamefont {Barrett}}]{Arnold2024b}%
  \BibitemOpen
  \bibfield  {author} {\bibinfo {author} {\bibfnamefont {K.~J.}\ \bibnamefont {Arnold}}, \bibinfo {author} {\bibfnamefont {S.}~\bibnamefont {Bustabad}}, \bibinfo {author} {\bibfnamefont {Z.}~\bibnamefont {Qi}}, \bibinfo {author} {\bibfnamefont {Q.}~\bibnamefont {Qichen}}, \bibinfo {author} {\bibfnamefont {Z.}~\bibnamefont {Zhang}}, \bibinfo {author} {\bibfnamefont {Z.}~\bibnamefont {Zhao}},\ and\ \bibinfo {author} {\bibfnamefont {M.~D.}\ \bibnamefont {Barrett}},\ }\bibfield  {title} {\bibinfo {title} {Validating a lutetium frequency reference.},\ }\href {https://doi.org/10.1088/1742-6596/2889/1/012040} {\bibfield  {journal} {\bibinfo  {journal} {J. Phys. Conf. Ser.}\ }\textbf {\bibinfo {volume} {2889}},\ \bibinfo {pages} {012040} (\bibinfo {year} {2024})} \BibitemShut {NoStop}%
\bibitem [{\citenamefont {Lindvall}\ \emph {et~al.}(2022)\citenamefont {Lindvall}, \citenamefont {Hanhij\"arvi}, \citenamefont {Fordell},\ and\ \citenamefont {Wallin}}]{Lindvall2022a}%
  \BibitemOpen
  \bibfield  {author} {\bibinfo {author} {\bibfnamefont {T.}~\bibnamefont {Lindvall}}, \bibinfo {author} {\bibfnamefont {K.~J.}\ \bibnamefont {Hanhij\"arvi}}, \bibinfo {author} {\bibfnamefont {T.}~\bibnamefont {Fordell}},\ and\ \bibinfo {author} {\bibfnamefont {A.~E.}\ \bibnamefont {Wallin}},\ }\bibfield  {title} {\bibinfo {title} {High-accuracy determination of {P}aul-trap stability parameters for electric-quadrupole-shift prediction},\ }\href {https://doi.org/10.1063/5.0106633} {\bibfield  {journal} {\bibinfo  {journal} {J. Appl. Phys.}\ }\textbf {\bibinfo {volume} {132}},\ \bibinfo {pages} {124401} (\bibinfo {year} {2022})}\BibitemShut {NoStop}%
\bibitem [{\citenamefont {Keller}\ \emph {et~al.}(2015)\citenamefont {Keller}, \citenamefont {Partner}, \citenamefont {Burgermeister},\ and\ \citenamefont {Mehlst\"aubler}}]{Keller2015a}%
  \BibitemOpen
  \bibfield  {author} {\bibinfo {author} {\bibfnamefont {J.}~\bibnamefont {Keller}}, \bibinfo {author} {\bibfnamefont {H.~L.}\ \bibnamefont {Partner}}, \bibinfo {author} {\bibfnamefont {T.}~\bibnamefont {Burgermeister}},\ and\ \bibinfo {author} {\bibfnamefont {T.~E.}\ \bibnamefont {Mehlst\"aubler}},\ }\bibfield  {title} {\bibinfo {title} {{P}recise determination of micromotion for trapped-ion optical clocks},\ }\href {https://doi.org/10.1063/1.4930037} {\bibfield  {journal} {\bibinfo  {journal} {J. Appl. Phys.}\ }\textbf {\bibinfo {volume} {118}},\ \bibinfo {eid} {104501} (\bibinfo {year} {2015})}\BibitemShut {NoStop}%
\bibitem [{Note1()}]{Note1}%
  \BibitemOpen
  \bibinfo {note} {See Supplemental Material [URL will be inserted by publisher], which includes Ref.~\cite {Dube2014a,Lindvall2022a,Schrama1993a,Bentine2020a}, for derivation of equations and treatment of trap anharmonicity.}\BibitemShut {Stop}%
\bibitem [{\citenamefont {Schrama}\ \emph {et~al.}(1993)\citenamefont {Schrama}, \citenamefont {Peik}, \citenamefont {Smith},\ and\ \citenamefont {Walther}}]{Schrama1993a}%
  \BibitemOpen
  \bibfield  {author} {\bibinfo {author} {\bibfnamefont {C.~A.}\ \bibnamefont {Schrama}}, \bibinfo {author} {\bibfnamefont {E.}~\bibnamefont {Peik}}, \bibinfo {author} {\bibfnamefont {W.~W.}\ \bibnamefont {Smith}},\ and\ \bibinfo {author} {\bibfnamefont {H.}~\bibnamefont {Walther}},\ }\bibfield  {title} {\bibinfo {title} {{N}ovel miniature ion traps},\ }\href {https://doi.org/10.1016/0030-4018(93)90318-Y} {\bibfield  {journal} {\bibinfo  {journal} {Opt. Commun.}\ }\textbf {\bibinfo {volume} {101}},\ \bibinfo {pages} {32} (\bibinfo {year} {1993})}\BibitemShut {NoStop}%
\bibitem [{\citenamefont {Bentine}\ \emph {et~al.}(2020)\citenamefont {Bentine}, \citenamefont {Foot},\ and\ \citenamefont {Trypogeorgos}}]{Bentine2020a}%
  \BibitemOpen
  \bibfield  {author} {\bibinfo {author} {\bibfnamefont {E.}~\bibnamefont {Bentine}}, \bibinfo {author} {\bibfnamefont {C.}~\bibnamefont {Foot}},\ and\ \bibinfo {author} {\bibfnamefont {D.}~\bibnamefont {Trypogeorgos}},\ }\bibfield  {title} {\bibinfo {title} {{(py)LIon}: A package for simulating trapped ion trajectories},\ }\href {https://doi.org/https://doi.org/10.1016/j.cpc.2020.107187} {\bibfield  {journal} {\bibinfo  {journal} {Comput. Phys. Commun.}\ }\textbf {\bibinfo {volume} {253}},\ \bibinfo {pages} {107187} (\bibinfo {year} {2020})}\BibitemShut {NoStop}%
\bibitem [{Note2()}]{Note2}%
  \BibitemOpen
  \bibinfo {note} {Note that in \cite {Dube2014a,Huang2019a}, the angle $\beta $ is the angle between the total rf electric field $\protect \mathbf {E}$ and the trap $z$ axis.}\BibitemShut {Stop}%
\bibitem [{\citenamefont {Dub\'e}\ \emph {et~al.}(2005)\citenamefont {Dub\'e}, \citenamefont {Madej}, \citenamefont {Bernard}, \citenamefont {Marmet}, \citenamefont {Boulanger},\ and\ \citenamefont {Cundy}}]{Dube2005a}%
  \BibitemOpen
  \bibfield  {author} {\bibinfo {author} {\bibfnamefont {P.}~\bibnamefont {Dub\'e}}, \bibinfo {author} {\bibfnamefont {A.~A.}\ \bibnamefont {Madej}}, \bibinfo {author} {\bibfnamefont {J.~E.}\ \bibnamefont {Bernard}}, \bibinfo {author} {\bibfnamefont {L.}~\bibnamefont {Marmet}}, \bibinfo {author} {\bibfnamefont {J.-S.}\ \bibnamefont {Boulanger}},\ and\ \bibinfo {author} {\bibfnamefont {S.}~\bibnamefont {Cundy}},\ }\bibfield  {title} {\bibinfo {title} {{E}lectric {Q}uadrupole {S}hift {C}ancellation in {S}ingle-{I}on {O}ptical {F}requency {S}tandards},\ }\href {https://doi.org/10.1103/PhysRevLett.95.033001} {\bibfield  {journal} {\bibinfo  {journal} {Phys. Rev. Lett.}\ }\textbf {\bibinfo {volume} {95}},\ \bibinfo {pages} {033001} (\bibinfo {year} {2005})}\BibitemShut {NoStop}%
\bibitem [{Note3()}]{Note3}%
  \BibitemOpen
  \bibinfo {note} {This requires the azimuthal angle of the rf direction, $\varphi _\Omega $, whose uncertainty of ${\approx }5^\circ $ has a completely negligible effect.}\BibitemShut {Stop}%
\bibitem [{\citenamefont {Jiang}\ \emph {et~al.}(2009)\citenamefont {Jiang}, \citenamefont {Arora}, \citenamefont {Safronova},\ and\ \citenamefont {Clark}}]{Jiang2009a}%
  \BibitemOpen
  \bibfield  {author} {\bibinfo {author} {\bibfnamefont {D.}~\bibnamefont {Jiang}}, \bibinfo {author} {\bibfnamefont {B.}~\bibnamefont {Arora}}, \bibinfo {author} {\bibfnamefont {M.~S.}\ \bibnamefont {Safronova}},\ and\ \bibinfo {author} {\bibfnamefont {C.~W.}\ \bibnamefont {Clark}},\ }\bibfield  {title} {\bibinfo {title} {{B}lackbody-radiation shift in a $^{88}${S}r$^+$ ion optical frequency standard},\ }\href {https://doi.org/10.1088/0953-4075/42/15/154020} {\bibfield  {journal} {\bibinfo  {journal} {J. Phys. B: At. Mol. Opt. Phys.}\ }\textbf {\bibinfo {volume} {42}},\ \bibinfo {pages} {154020} (\bibinfo {year} {2009})}\BibitemShut {NoStop}%
\bibitem [{\citenamefont {Madej}\ \emph {et~al.}(2004)\citenamefont {Madej}, \citenamefont {Bernard}, \citenamefont {Dub\'e}, \citenamefont {Marmet},\ and\ \citenamefont {Windeler}}]{Madej2004a}%
  \BibitemOpen
  \bibfield  {author} {\bibinfo {author} {\bibfnamefont {A.~A.}\ \bibnamefont {Madej}}, \bibinfo {author} {\bibfnamefont {J.~E.}\ \bibnamefont {Bernard}}, \bibinfo {author} {\bibfnamefont {P.}~\bibnamefont {Dub\'e}}, \bibinfo {author} {\bibfnamefont {L.}~\bibnamefont {Marmet}},\ and\ \bibinfo {author} {\bibfnamefont {R.~S.}\ \bibnamefont {Windeler}},\ }\bibfield  {title} {\bibinfo {title} {{A}bsolute frequency of the $^{88}${S}r$^+$ $5s \; ^2{S}_{1/2}$ -- $4d \; ^2{D}_{5/2}$ reference transition at 445 {TH}z and evaluation of systematic shifts},\ }\href {https://doi.org/10.1103/PhysRevA.70.012507} {\bibfield  {journal} {\bibinfo  {journal} {Phys. Rev. A}\ }\textbf {\bibinfo {volume} {70}},\ \bibinfo {pages} {012507} (\bibinfo {year} {2004})}\BibitemShut {NoStop}%
\bibitem [{\citenamefont {Shiner}\ \emph {et~al.}(2007)\citenamefont {Shiner}, \citenamefont {Madej}, \citenamefont {Dub\'e},\ and\ \citenamefont {Bernard}}]{Shiner2007a}%
  \BibitemOpen
  \bibfield  {author} {\bibinfo {author} {\bibfnamefont {A.~D.}\ \bibnamefont {Shiner}}, \bibinfo {author} {\bibfnamefont {A.~A.}\ \bibnamefont {Madej}}, \bibinfo {author} {\bibfnamefont {P.}~\bibnamefont {Dub\'e}},\ and\ \bibinfo {author} {\bibfnamefont {J.~E.}\ \bibnamefont {Bernard}},\ }\bibfield  {title} {\bibinfo {title} {{A}bsolute optical frequency measurement of saturated absorption lines in {R}b near 422 nm},\ }\href {https://doi.org/10.1007/s00340-007-2836-y} {\bibfield  {journal} {\bibinfo  {journal} {Appl. Phys. B}\ }\textbf {\bibinfo {volume} {89}},\ \bibinfo {pages} {595} (\bibinfo {year} {2007})}\BibitemShut {NoStop}%
\bibitem [{\citenamefont {Lindvall}\ \emph {et~al.}(2013)\citenamefont {Lindvall}, \citenamefont {Fordell}, \citenamefont {Tittonen},\ and\ \citenamefont {Merimaa}}]{Lindvall2013a}%
  \BibitemOpen
  \bibfield  {author} {\bibinfo {author} {\bibfnamefont {T.}~\bibnamefont {Lindvall}}, \bibinfo {author} {\bibfnamefont {T.}~\bibnamefont {Fordell}}, \bibinfo {author} {\bibfnamefont {I.}~\bibnamefont {Tittonen}},\ and\ \bibinfo {author} {\bibfnamefont {M.}~\bibnamefont {Merimaa}},\ }\bibfield  {title} {\bibinfo {title} {{U}npolarized, incoherent repumping light for prevention of dark states in a trapped and laser-cooled single ion},\ }\href {https://doi.org/10.1103/PhysRevA.87.013439} {\bibfield  {journal} {\bibinfo  {journal} {Phys. Rev. A}\ }\textbf {\bibinfo {volume} {87}},\ \bibinfo {pages} {013439} (\bibinfo {year} {2013})}\BibitemShut {NoStop}%
\bibitem [{\citenamefont {Fordell}\ \emph {et~al.}(2015)\citenamefont {Fordell}, \citenamefont {Lindvall}, \citenamefont {Dub\'{e}}, \citenamefont {Madej}, \citenamefont {Wallin},\ and\ \citenamefont {Merimaa}}]{Fordell2015a}%
  \BibitemOpen
  \bibfield  {author} {\bibinfo {author} {\bibfnamefont {T.}~\bibnamefont {Fordell}}, \bibinfo {author} {\bibfnamefont {T.}~\bibnamefont {Lindvall}}, \bibinfo {author} {\bibfnamefont {P.}~\bibnamefont {Dub\'{e}}}, \bibinfo {author} {\bibfnamefont {A.~A.}\ \bibnamefont {Madej}}, \bibinfo {author} {\bibfnamefont {A.~E.}\ \bibnamefont {Wallin}},\ and\ \bibinfo {author} {\bibfnamefont {M.}~\bibnamefont {Merimaa}},\ }\bibfield  {title} {\bibinfo {title} {{B}roadband, unpolarized repumping and clearout light sources for {S}r$^+$ single-ion clocks},\ }\href {https://doi.org/10.1364/OL.40.001822} {\bibfield  {journal} {\bibinfo  {journal} {Opt. Lett.}\ }\textbf {\bibinfo {volume} {40}},\ \bibinfo {pages} {1822} (\bibinfo {year} {2015})}\BibitemShut {NoStop}%
\bibitem [{\citenamefont {H\"afner}\ \emph {et~al.}(2015)\citenamefont {H\"afner}, \citenamefont {Falke}, \citenamefont {Grebing}, \citenamefont {Vogt}, \citenamefont {Legero}, \citenamefont {Merimaa}, \citenamefont {Lisdat},\ and\ \citenamefont {Sterr}}]{Hafner2015a}%
  \BibitemOpen
  \bibfield  {author} {\bibinfo {author} {\bibfnamefont {S.}~\bibnamefont {H\"afner}}, \bibinfo {author} {\bibfnamefont {S.}~\bibnamefont {Falke}}, \bibinfo {author} {\bibfnamefont {C.}~\bibnamefont {Grebing}}, \bibinfo {author} {\bibfnamefont {S.}~\bibnamefont {Vogt}}, \bibinfo {author} {\bibfnamefont {T.}~\bibnamefont {Legero}}, \bibinfo {author} {\bibfnamefont {M.}~\bibnamefont {Merimaa}}, \bibinfo {author} {\bibfnamefont {C.}~\bibnamefont {Lisdat}},\ and\ \bibinfo {author} {\bibfnamefont {U.}~\bibnamefont {Sterr}},\ }\bibfield  {title} {\bibinfo {title} {$8 \times 10^{-17}$ fractional laser frequency instability with a long room-temperature cavity},\ }\href {https://doi.org/10.1364/OL.40.002112} {\bibfield  {journal} {\bibinfo  {journal} {Opt. Lett.}\ }\textbf {\bibinfo {volume} {40}},\ \bibinfo {pages} {2112} (\bibinfo {year} {2015})}\BibitemShut {NoStop}%
\bibitem [{\citenamefont {Lindvall}\ \emph {et~al.}(2023)\citenamefont {Lindvall}, \citenamefont {Wallin}, \citenamefont {Hanhij\"arvi},\ and\ \citenamefont {Fordell}}]{Lindvall2023a}%
  \BibitemOpen
  \bibfield  {author} {\bibinfo {author} {\bibfnamefont {T.}~\bibnamefont {Lindvall}}, \bibinfo {author} {\bibfnamefont {A.~E.}\ \bibnamefont {Wallin}}, \bibinfo {author} {\bibfnamefont {K.~J.}\ \bibnamefont {Hanhij\"arvi}},\ and\ \bibinfo {author} {\bibfnamefont {T.}~\bibnamefont {Fordell}},\ }\bibfield  {title} {\bibinfo {title} {Noise-induced servo errors in optical clocks utilizing {R}abi interrogation},\ }\href {https://doi.org/10.1088/1681-7575/acdfd4} {\bibfield  {journal} {\bibinfo  {journal} {Metrologia}\ }\textbf {\bibinfo {volume} {60}},\ \bibinfo {pages} {045008} (\bibinfo {year} {2023})}\BibitemShut {NoStop}%
\bibitem [{Note4()}]{Note4}%
  \BibitemOpen
  \bibinfo {note} {T.~Lindvall, K.~J.~Hanhij\"arvi, T.~Fordell, and A.~E.~Wallin, Zenodo, 2025, \href{http://doi.org/10.5281/zenodo.15793152}{http://doi.org/10.5281/zenodo.15793152}.}\BibitemShut {Stop}%
\end{thebibliography}

\begin{thebibliography}{4}%
\makeatletter
\providecommand \@ifxundefined [1]{%
 \@ifx{#1\undefined}
}%
\providecommand \@ifnum [1]{%
 \ifnum #1\expandafter \@firstoftwo
 \else \expandafter \@secondoftwo
 \fi
}%
\providecommand \@ifx [1]{%
 \ifx #1\expandafter \@firstoftwo
 \else \expandafter \@secondoftwo
 \fi
}%
\providecommand \natexlab [1]{#1}%
\providecommand \enquote  [1]{``#1''}%
\providecommand \bibnamefont  [1]{#1}%
\providecommand \bibfnamefont [1]{#1}%
\providecommand \citenamefont [1]{#1}%
\providecommand \href@noop [0]{\@secondoftwo}%
\providecommand \href [0]{\begingroup \@sanitize@url \@href}%
\providecommand \@href[1]{\@@startlink{#1}\@@href}%
\providecommand \@@href[1]{\endgroup#1\@@endlink}%
\providecommand \@sanitize@url [0]{\catcode `\\12\catcode `\$12\catcode `\&12\catcode `\#12\catcode `\^12\catcode `\_12\catcode `\%12\relax}%
\providecommand \@@startlink[1]{}%
\providecommand \@@endlink[0]{}%
\providecommand \url  [0]{\begingroup\@sanitize@url \@url }%
\providecommand \@url [1]{\endgroup\@href {#1}{\urlprefix }}%
\providecommand \urlprefix  [0]{URL }%
\providecommand \Eprint [0]{\href }%
\providecommand \doibase [0]{https://doi.org/}%
\providecommand \selectlanguage [0]{\@gobble}%
\providecommand \bibinfo  [0]{\@secondoftwo}%
\providecommand \bibfield  [0]{\@secondoftwo}%
\providecommand \translation [1]{[#1]}%
\providecommand \BibitemOpen [0]{}%
\providecommand \bibitemStop [0]{}%
\providecommand \bibitemNoStop [0]{.\EOS\space}%
\providecommand \EOS [0]{\spacefactor3000\relax}%
\providecommand \BibitemShut  [1]{\csname bibitem#1\endcsname}%
\let\auto@bib@innerbib\@empty
\bibitem [{\citenamefont {Dub\'e}\ \emph {et~al.}(2014)\citenamefont {Dub\'e}, \citenamefont {Madej}, \citenamefont {Tibbo},\ and\ \citenamefont {Bernard}}]{sDube2014a}%
  \BibitemOpen
  \bibfield  {author} {\bibinfo {author} {\bibfnamefont {P.}~\bibnamefont {Dub\'e}}, \bibinfo {author} {\bibfnamefont {A.~A.}\ \bibnamefont {Madej}}, \bibinfo {author} {\bibfnamefont {M.}~\bibnamefont {Tibbo}},\ and\ \bibinfo {author} {\bibfnamefont {J.~E.}\ \bibnamefont {Bernard}},\ }\bibfield  {title} {\bibinfo {title} {{H}igh-{A}ccuracy {M}easurement of the {D}ifferential {S}calar {P}olarizability of a $^{88}${Sr}$^{+}$ {C}lock {U}sing the {T}ime-{D}ilation {E}ffect},\ }\href {https://doi.org/10.1103/PhysRevLett.112.173002} {\bibfield  {journal} {\bibinfo  {journal} {Phys. Rev. Lett.}\ }\textbf {\bibinfo {volume} {112}},\ \bibinfo {pages} {173002} (\bibinfo {year} {2014})}\BibitemShut {NoStop}%
\bibitem [{\citenamefont {Lindvall}\ \emph {et~al.}(2022)\citenamefont {Lindvall}, \citenamefont {Hanhij\"arvi}, \citenamefont {Fordell},\ and\ \citenamefont {Wallin}}]{sLindvall2022a}%
  \BibitemOpen
  \bibfield  {author} {\bibinfo {author} {\bibfnamefont {T.}~\bibnamefont {Lindvall}}, \bibinfo {author} {\bibfnamefont {K.~J.}\ \bibnamefont {Hanhij\"arvi}}, \bibinfo {author} {\bibfnamefont {T.}~\bibnamefont {Fordell}},\ and\ \bibinfo {author} {\bibfnamefont {A.~E.}\ \bibnamefont {Wallin}},\ }\bibfield  {title} {\bibinfo {title} {High-accuracy determination of {P}aul-trap stability parameters for electric-quadrupole-shift prediction},\ }\href {https://doi.org/10.1063/5.0106633} {\bibfield  {journal} {\bibinfo  {journal} {J. Appl. Phys.}\ }\textbf {\bibinfo {volume} {132}},\ \bibinfo {pages} {124401} (\bibinfo {year} {2022})}\BibitemShut {NoStop}%
\bibitem [{\citenamefont {Schrama}\ \emph {et~al.}(1993)\citenamefont {Schrama}, \citenamefont {Peik}, \citenamefont {Smith},\ and\ \citenamefont {Walther}}]{sSchrama1993a}%
  \BibitemOpen
  \bibfield  {author} {\bibinfo {author} {\bibfnamefont {C.~A.}\ \bibnamefont {Schrama}}, \bibinfo {author} {\bibfnamefont {E.}~\bibnamefont {Peik}}, \bibinfo {author} {\bibfnamefont {W.~W.}\ \bibnamefont {Smith}},\ and\ \bibinfo {author} {\bibfnamefont {H.}~\bibnamefont {Walther}},\ }\bibfield  {title} {\bibinfo {title} {{N}ovel miniature ion traps},\ }\href {https://doi.org/10.1016/0030-4018(93)90318-Y} {\bibfield  {journal} {\bibinfo  {journal} {Opt. Commun.}\ }\textbf {\bibinfo {volume} {101}},\ \bibinfo {pages} {32} (\bibinfo {year} {1993})}\BibitemShut {NoStop}%
\bibitem [{\citenamefont {Bentine}\ \emph {et~al.}(2020)\citenamefont {Bentine}, \citenamefont {Foot},\ and\ \citenamefont {Trypogeorgos}}]{sBentine2020a}%
  \BibitemOpen
  \bibfield  {author} {\bibinfo {author} {\bibfnamefont {E.}~\bibnamefont {Bentine}}, \bibinfo {author} {\bibfnamefont {C.}~\bibnamefont {Foot}},\ and\ \bibinfo {author} {\bibfnamefont {D.}~\bibnamefont {Trypogeorgos}},\ }\bibfield  {title} {\bibinfo {title} {{(py)LIon}: A package for simulating trapped ion trajectories},\ }\href {https://doi.org/https://doi.org/10.1016/j.cpc.2020.107187} {\bibfield  {journal} {\bibinfo  {journal} {Comput. Phys. Commun.}\ }\textbf {\bibinfo {volume} {253}},\ \bibinfo {pages} {107187} (\bibinfo {year} {2020})}\BibitemShut {NoStop}%
\end{thebibliography}
\end{document}